%% file: main.tex
\documentclass[10pt,journal]{IEEEtran}
\usepackage{array}
\usepackage{amsmath,amssymb,amsfonts}
\usepackage{booktabs}
\usepackage{balance}
\usepackage{etoolbox}

\makeatletter
\patchcmd{\@BAlancecol}{\setbox\@leftcolumn}
  {\advance\@tempdima by 2pt\setbox\@leftcolumn}{}
  {\PackageError{manuscript}{Unable to adjust bibliography balancing}{Check the balance package version.}}
\makeatother
\usepackage{cite}
\usepackage{graphicx}
\usepackage{makecell}
\usepackage{multirow}
\usepackage{placeins}
\usepackage[caption=false,font=footnotesize]{subfig}
\usepackage{tikz}
\usetikzlibrary{arrows.meta,positioning,fit,shapes.geometric,calc}
\usepackage{url}
\usepackage{xcolor}

\graphicspath{{figures/}}
\newcolumntype{P}[1]{>{\raggedright\arraybackslash}p{#1}}

\begin{document}

\title{Integrated Sensing and Communication in 3GPP: Standardization, Performance Evaluation, and Experimental Validation Toward 6G}

\author{Gosan Noh, Gebremeskel Seiley Abraha, Joshua Yabana, Nathaniel Foli, Michael Ankomah Asare,\\
Evelyn Seyram Kakah, Mildred Nyantakyiwaa Okyere, De-Roy Ekow Sagoe, Shrayan Das, Pekka Ky\"osti,\\
Junhyeong Kim, Kyoung-Jae Lee, and Eunkyung Kim%
\thanks{G. Noh, J. Yabana, N. Foli, M. A. Asare, E. S. Kakah, M. N. Okyere, and D.-R. E. Sagoe are with the Department of Electronic Engineering, Hanbat National University, Daejeon 34158, South Korea.}%
\thanks{G. S. Abraha and K.-J. Lee are with the School of Electrical and Electronics Engineering, Chung-Ang University, Seoul 06974, South Korea.}%
\thanks{S. Das and P. Ky\"osti are with the Centre for Wireless Communications, University of Oulu, 90014 Oulu, Finland. P. Ky\"osti is also with Keysight Technologies Finland Oy, 90630 Oulu, Finland.}%
\thanks{J. Kim is with 6G Wireless Access System Research Section, Electronics and Telecommunications Research Institute (ETRI), Daejeon 34129, South Korea. He is also with the Major in Information and Communication Engineering, ETRI School, University of Science and Technology (UST), Daejeon 34113, South Korea}%
\thanks{E. Kim is with the Department of Artificial Intelligence Software, Hanbat National University, Sejong 30139, South Korea.}%
\thanks{This work was supported by the Institute of Information \& Communications Technology Planning \& Evaluation (IITP) grant funded by the Korean government (MSIT) (No. RS-2024-00444230, Development of Wireless Technology for Integrated Sensing and Communication). This work was also supported in part by the IITP-ITRC (Information Technology Research Center) grant funded by the Korean government (Ministry of Science and ICT) (IITP-2025-RS-2024-00437886). This work was partly supported by the 6G-WISECOM project (Grant no. 3630/31/2024) funded by Business Finland, and partly by the Research Council of Finland 6G Flagship (Grant no. 369116).}}

\maketitle

\begin{abstract}
Integrated sensing and communication (ISAC) is evolving from a primarily physical-layer research into a standardized cellular-network capability spanning channel modeling, radio procedures, network architecture, and the delivery of sensing services. We review this progression within the Third Generation Partnership Project (3GPP), from Release~19 service requirements and target-aware channel modeling to the Release~20 New Radio (NR) sensing evaluation framework, radio access network procedures, and system architecture, and discuss future directions toward 6G. We then present two 3GPP-aligned simulation case studies to demonstrate the corresponding evaluation approaches. The first case study compares hard-decision, soft-decision, and hybrid fusion for unmanned aerial vehicle (UAV) sensing using multiple transmission and reception points (TRPs). Detection, false alarms, and conditional position and velocity accuracy are evaluated using receiver-derived angle estimates and separately calibrated method-specific thresholds. The second case study compares ideal and practical sensing-assisted predictive handover with conventional handover, while accounting for sensing error and information age. We further present a testbed-based measurement study that characterizes monostatic and bistatic UAV radar cross sections across frequencies, platforms, and observation angles and provides empirical support for correlated target-response modeling. Finally, we discuss open challenges and future research directions in multi-node fusion, resource and interference management, network evolution, security, privacy, and reproducible evaluation. Together, these components connect calibrated channel models, link- and system-level simulation, real-time prototypes, controlled measurements, and field trials through a staged evaluation roadmap.
\end{abstract}


\input{sections/01_introduction}
\input{sections/02_review_methodology}
\input{sections/03_foundations}

\input{sections/04_standardization}
\input{sections/05_simulation}
\input{sections/06_testbeds}
\input{sections/07_challenges}
\input{sections/08_conclusion}

\IEEEtriggercmd{\balance}
\IEEEtriggeratref{71}
\bibliographystyle{IEEEtran}
\bibliography{IEEEabrv,references}

\end{document}

%% file: sections/01_introduction.tex
\section{Introduction}
\label{sec:introduction}

Wireless networks are evolving from communication-centric infrastructures into systems that can also perceive physical objects and their environments. Integrated sensing and communication (ISAC) supports this evolution by reusing radio spectrum, waveforms, antenna arrays, and network infrastructure for data delivery and sensing~\cite{liu_2023_spm,liu_2022_jsac}. The International Telecommunication Union identifies ISAC as one of the usage scenarios in its framework for International Mobile Telecommunications for 2030 and beyond (IMT-2030)~\cite{itur_m2160}. Representative ISAC applications include unmanned aerial vehicle (UAV) monitoring, transportation safety, industrial automation, environmental awareness, and digital twins. Sensing information can also support beam management, blockage prediction, and mobility support, while cellular connectivity provides synchronization, coordination, and wide-area information exchange for distributed sensing~\cite{zhang_2022_cst,gonzalez_2024_pieee}. Hence, practical ISAC realization extends beyond physical-layer waveform design and requires consistent channel models, evaluation frameworks, radio procedures, network functions, and service provisioning mechanisms.

The Third Generation Partnership Project (3GPP) has progressively developed these foundational elements. Release~19 established potential use cases and incorporated ISAC channel modeling into TR~38.901, including explicit sensing targets, separate target and background channels, radar cross section (RCS), bistatic geometry, and target-induced delay and Doppler~\cite{tr_22_837,ref11}. Release~20 extends this foundation toward New Radio (NR) ISAC feasibility evaluation and a deployable architecture. The initial radio baseline focuses on sensing UAV targets using an NR base station (gNB). In parallel, Radio Access Network (RAN) Working Group~3 (RAN3) studies sensing control and measurement transfer between the gNB and the Sensing Function (SenF), while Service and System Aspects (SA) Working Group~2 (SA2) develops service-level sensing management and sensing-service provisioning~\cite{tr38765,tr2370014,ts23137}. These activities need to be considered jointly across RAN Working Group~1 (RAN1) evaluation, RAN3 realization, and SA2 service coordination.

\begin{figure}[!t]
\centering
\includegraphics[width=0.9\columnwidth]{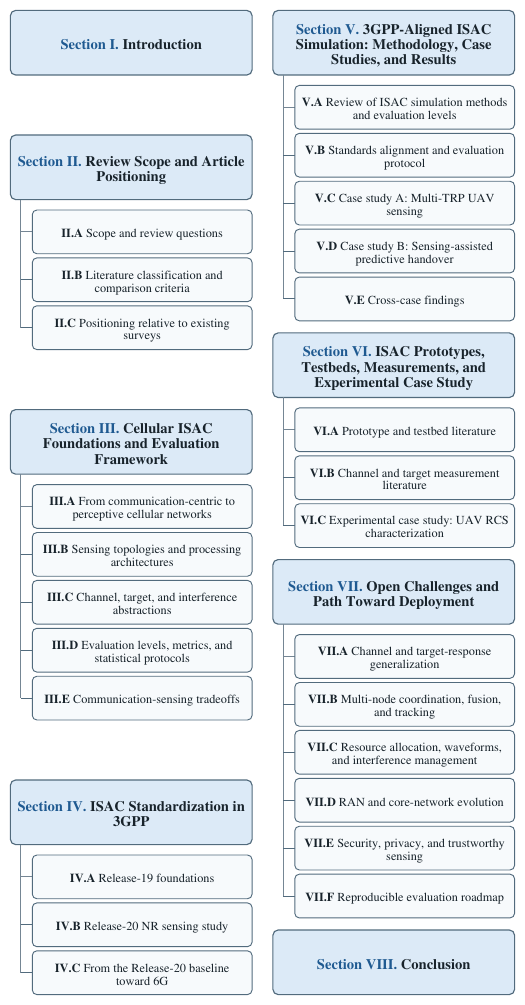}
\caption{Structure of this article. Sections~I--IV establish the review scope, technical foundations, and relevant 3GPP developments. Sections~V and VI present the simulation and experimental case studies, while Section~VII summarizes the open challenges and evaluation roadmap.}
\label{fig:paper_structure}
\end{figure}

Existing ISAC surveys provide comprehensive treatments of signal processing,
waveform design, resource allocation, channel and target modeling,
system-level simulation, architectures, prototypes, and testbeds
~\cite{liu_2022_jsac,gonzalez_2024_pieee,wei2025isacchannel,
luo2026isacsurvey,pedersen2024isacsls}. A remaining question is how these
technical elements relate to the assumptions and performance boundaries of
cellular standardization. We address this question by connecting the evolving 3GPP channel, radio, RAN, and system-architecture work with standards-aligned simulation case studies and measurement-based target characterization. More specifically, the main contributions are summarized as follows.

\begin{itemize}
\item \textbf{Cross-layer 3GPP framework:}
The article provides a unified interpretation of the 3GPP ISAC development path, connecting Release~19 service requirements and channel modeling with the Release~20 RAN1 evaluation framework, RAN3 sensing procedures, and SA2 service architecture, and discusses their coordinated evolution toward 6G.

\item \textbf{Evaluation taxonomy and simulation case studies:}
The article establishes a taxonomy of ISAC evaluation methodologies according to their modeling scope, implementation fidelity, and validation basis. Building on this taxonomy, UAV sensing with multiple transmission and reception points (TRPs) and sensing-assisted predictive handover are investigated through standards-aligned case studies that explicitly consider operating-point calibration, sensing uncertainty, conditional performance metrics, and the validity limits of statistical conclusions.

\item \textbf{Experimental validation and measurement-based characterization:}
The article organizes representative ISAC prototypes, network demonstrations, and measurement campaigns according to their functions and validation objectives. A UAV radar-cross-section measurement study further demonstrates how frequency, target aspect, and sensing geometry affect target response and should be incorporated into channel modeling and system evaluation.

\item \textbf{Standards-to-deployment roadmap:}
The article presents an evaluation roadmap that relates standardized scenarios and requirements to calibrated channel and link models, system-level evaluation, real-time prototypes, controlled measurements, and field trials. Based on this progression, it identifies research priorities in multi-node fusion, coexistence, network evolution, security, privacy, and reproducible evaluation.
\end{itemize}

Fig.~\ref{fig:paper_structure} summarizes the article organization and its progression from standards to deployment. Section~\ref{sec:review_methodology} defines the review scope, validation categories, and the article's distinctive positioning relative to existing surveys. Section~\ref{sec:foundations} introduces cellular ISAC configurations, channel models, performance metrics, and evaluation methodologies. Section~\ref{sec:standardization} traces the corresponding 3GPP development from Release~19 channel and service studies to the Release~20 radio and system-level work and its evolution toward 6G. Sections~\ref{sec:simulation} and~\ref{sec:testbeds} present standards-aligned simulation and measurement case studies, respectively, supported by corresponding reviews of the relevant literature. Section~\ref{sec:challenges} develops the open challenges and standards-to-deployment roadmap, and Section~\ref{sec:conclusion} concludes the article.

%% file: sections/02_review_methodology.tex
\section{Review Scope and Article Positioning}
\label{sec:review_methodology}

This section defines the scope and organization of the review and clarifies its distinction from existing ISAC surveys. The considered literature spans cellular standardization, channel and system evaluation, prototype implementation, and measurement-based validation. Studies are organized according to their technical scope and validation approach to establish a common basis for the subsequent standardization, simulation, and experimental discussions.

\subsection{Scope and Review Questions}

A central question guiding this review is: What technical foundations and validation steps are required to progress from a cellular sensing requirement to an implementable ISAC function? Answering this question requires more than a review of waveforms or sensing algorithms. The target characteristics and propagation channel determine what can be observed, while the radio configuration determines how the observations are acquired. The RAN and core-network functions govern how sensing is controlled and delivered, and the evaluation methodology determines how the resulting performance findings can be interpreted. We therefore examine these elements as parts of an integrated evaluation framework rather than as independent research topics.

Accordingly, the review addresses three questions. First, which ISAC functions, scenarios, channel models, metrics, and interfaces have been specified or studied by 3GPP? Second, which link-level, system-level, and multi-node simulation abstractions are required to evaluate the performance of sensing and sensing-assisted communication without attributing an operating-point change to an algorithmic gain? Third, which prototype, testbed, channel, and target measurements are needed to validate the assumptions used by those simulations? The review primarily focuses on cellular ISAC but also considers earlier research on joint radar and communication systems that establishes models, metrics, or implementation principles relevant to 5G Advanced and 6G ISAC.

\subsection{Literature Classification and Comparison Criteria}

The review covers 3GPP standardization, RAN and core-network architectures and procedures, channel modeling and measurement, simulation methodology, and prototype and testbed implementation. We use 3GPP specifications and approved reports as the primary sources for established requirements, models, and procedures, while meeting contributions are considered only to clarify evolving assumptions or unresolved options. Academic results are interpreted according to their stated scenario, model, metric, and experimental configuration.

The literature is classified by evaluation level, sensing topology, deployment scenario, channel basis, implementation maturity, and reported key performance indicators. The reviewed material is organized into four complementary layers:
\begin{itemize}
    \item \textbf{Standards} define common assumptions, system requirements, procedures, and interoperable behavior.
    \item \textbf{Simulations} quantify performance under specified channel, deployment, and operating assumptions.
    \item \textbf{Prototypes and testbeds} assess implementation feasibility under controlled or operational conditions.
    \item \textbf{Measurements} evaluate the physical validity of propagation, channel, and target-response models.
\end{itemize}

These layers provide complementary forms of validation and should be
interpreted according to their respective scope.

For comparison, each study is examined in terms of its scenario, sensing topology, channel and target models, antenna configuration, operating point, evaluation level, implementation maturity, and performance metrics. Numerical results are compared directly only when the underlying assumptions and metric definitions are compatible. Otherwise, the comparison remains qualitative and accounts explicitly for differences in the adopted models and configurations.

\subsection{Positioning Relative to Existing Surveys}

Table~\ref{tab:survey_positioning} positions this article relative to the closest surveys and tutorials. Zhang \emph{et al.} established a mobile-network view of joint communication and radar sensing~\cite{zhang_2022_cst}, while Liu \emph{et al.} organized the corresponding fundamental limits~\cite{liu_2022_cst}. Wei \emph{et al.} concentrated on ISAC channel, target-RCS, and clutter modeling~\cite{wei2025isacchannel}. More recent surveys provided a layered treatment spanning technologies, standardization, prototypes, and testbeds~\cite{luo2026isacsurvey}, or organized ISAC through its evolution in spectrum, network architecture, sensing modality, and standards~\cite{zhang2026isacevolution}. Pedersen \emph{et al.} provided the complementary tutorial basis for 3GPP-oriented radio system-level simulation, although their scope is not specific to ISAC~\cite{pedersen2024isacsls}.

Recent work further clarifies the distinction among physical-layer, network-level, and standardization perspectives. Dai \emph{et al.} review multiple-input multiple-output (MIMO) orthogonal frequency-division multiplexing (OFDM) sensing models and parameter estimation in far- and near-field conditions~\cite{dai2026mimoofdmisac}. Armeniakos \emph{et al.} examine stochastic-geometry models for sensing and communication networks~\cite{armeniakos2026stochastic}, while Han \emph{et al.} review distributed cooperation, signaling, and synchronization~\cite{han2025network}. Lin's recent preprint directly addresses the Release~19/20 progression and its evolution toward 6G~\cite{lin2026isac3gpp}.

\begin{table*}[!t]
\centering
\caption{Positioning Relative to Closely Related Surveys and Tutorials}
\label{tab:survey_positioning}
\scriptsize
\setlength{\tabcolsep}{2.1pt}
\renewcommand{\arraystretch}{1.05}
\begin{tabular}{P{0.28\columnwidth}P{0.58\columnwidth}P{1\columnwidth}}
\toprule
\textbf{Work} & \textbf{Primary coverage} & \textbf{Relation to this article} \\
\midrule
Zhang \emph{et al.}~\cite{zhang_2022_cst} & Mobile-network sensing & Provides the early cellular and networked-sensing foundation \\
Liu \emph{et al.}~\cite{liu_2022_cst} & Fundamental limits & Provides the theoretical sensing--communication tradeoff foundation \\
Wei \emph{et al.}~\cite{wei2025isacchannel} & Channel, RCS, and clutter modeling & Provides the detailed channel and target-model taxonomy \\
Luo \emph{et al.}~\cite{luo2026isacsurvey} & Layered architecture, standards, and platforms & Provides broad technology, standardization, prototype, and testbed coverage \\
Zhang \emph{et al.}~\cite{zhang2026isacevolution} & Evolution of ISAC & Provides the RF/optical, network, sensing-modality, and multi-body standards evolution \\
Pedersen \emph{et al.}~\cite{pedersen2024isacsls} & 3GPP radio system-level simulation tutorial & Provides general simulation models, workflows, calibration, and software practices \\
Dai \emph{et al.}~\cite{dai2026mimoofdmisac} & MIMO-OFDM sensing and estimation & Provides far- and near-field signal models and processing methods \\
Armeniakos \emph{et al.}~\cite{armeniakos2026stochastic} & Stochastic-geometry evaluation & Provides spatial network models and sensing--communication performance analysis \\
Han \emph{et al.}~\cite{han2025network} & Network-level ISAC & Provides cooperation, distributed signaling, and synchronization design principles \\
Lin~\cite{lin2026isac3gpp} & 3GPP evolution from 5G-Advanced to 6G & Reviews Release~19/20 radio and architecture developments and future directions \\
\textbf{This article} & \textbf{Cellular standards-to-validation chain} & \textbf{Connects Release~19/20 SA1, RAN1, RAN3, and SA2 with two simulation cases and a UAV RCS measurement case} \\
\bottomrule
\end{tabular}
\end{table*}

This article connects the Release~19 channel model and Release~20 evaluation
assumptions with the sensing procedures and system architectures studied by
RAN3 and SA2. Two simulation case studies and a UAV target-response
measurement study illustrate how these standards assumptions can be translated into performance evaluation and experimental characterization.


%% file: sections/03_foundations.tex
\section{Cellular ISAC Foundations and Evaluation Framework}
\label{sec:foundations}

Cellular ISAC changes the role of the radio channel. A communication receiver treats propagation primarily as the medium through which data are conveyed, whereas a sensing receiver interprets delay, Doppler, angle, amplitude, and their temporal evolution as observations of physical objects. This section establishes common terminology and an evaluation framework that relates these physical observations to sensing outputs and network decisions. It distinguishes the \emph{propagation channel}, the \emph{target response}, the \emph{sensing algorithm}, and the \emph{network decision}, and clarifies how the assumptions at each level affect the reported performance.

\subsection{From Communication-Centric to Perceptive Cellular Networks}

\begin{figure}[!t]
\centering
\includegraphics[width=\columnwidth]{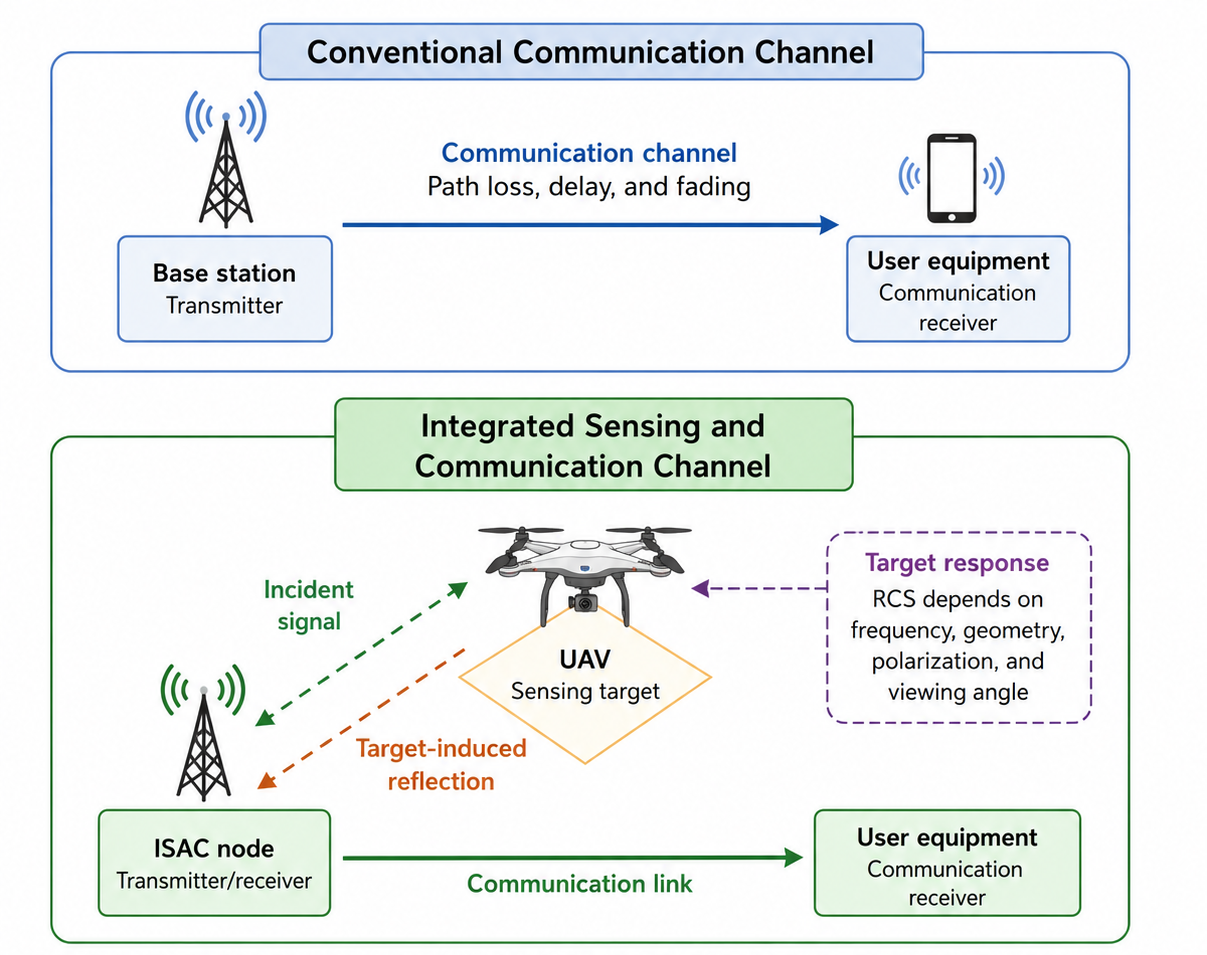}
\caption{Conceptual comparison of conventional communication and ISAC
channels. The target-induced path includes a UAV response that varies with
frequency, geometry, polarization, and sensing direction.}
\label{fig:isac_uav_concept}
\end{figure}

The advancement from distinct radar and communication platforms to integrated dual-functional radio networks is motivated by the potential to reuse spectrum, antenna arrays, radio-frequency hardware, synchronization mechanism, and existing network infrastructure~\cite{liu_2023_spm,liu_2022_jsac}. In a conventional cellular link, reflections, blockage, and mobility are represented mainly through channel gain, dispersion, and interference. In a perceptive network, the same effects also provide information about object location, motion, shape, or activity~\cite{zhang_2022_cst,gonzalez_2024_pieee}. As illustrated in Fig.~\ref{fig:isac_uav_concept}, the sensing channel therefore includes a target-induced contribution whose electromagnetic response must be distinguished from ordinary communication propagation. The practical objective extends beyond placing a radar estimator beside a communication receiver: the network must control how observations are acquired, combine them across nodes and time, and expose the resulting information with known accuracy, latency, and reliability.

Three functional relationships should be distinguished. First, \emph{joint sensing and communication} uses common or coordinated radio resources to produce both communication and sensing outputs. Second, \emph{sensing-assisted communication} relies on estimates of the environment or user motion to enhance beam management, blockage prediction, mobility control, scheduling, and resource allocation. Third, \emph{communication-assisted sensing} uses network synchronization, positioning, scheduling, transport, and distributed coordination to improve sensing. These relationships can coexist, but they represent different performance outcomes. For example, an accurate target-velocity estimate reflects sensing performance, whereas a reduction in handover interruption enabled by that estimate reflects network-control performance. The latter also depends on reporting delay, prediction error, policy thresholds, and radio-link behavior.

This bidirectional coupling creates a closed loop. Radio resources generate observations that sensing processing converts into measurements or tracks. Network control uses these outputs to make resource, beam, or mobility decisions, which subsequently change the observation geometry. Although this loop can improve network operation, stale or overconfident sensing information can degrade performance. A rigorous evaluation should therefore report both sensing quality and the resulting communication performance rather than assume that additional sensing information is always beneficial.

\subsection{Sensing Topologies and Processing Architectures}

A sensing topology is defined by the locations of the sensing transmitter (STx), sensing receiver (SRx), target, and processing function. The basic radio geometries are monostatic, bistatic, and multistatic. Their main implications are summarized in Table~\ref{tab:isac_topologies}. This classification is independent of whether the participating radio node is a TRP or a user equipment (UE). The corresponding 3GPP modes are introduced in Section~\ref{sec:standardization}.

The coherent processing interval (CPI) is the observation duration over which sensing samples are combined coherently.

\begin{table*}[!t]
\caption{Sensing topologies and their principal evaluation consequences}
\label{tab:isac_topologies}
\centering
\footnotesize
\setlength{\tabcolsep}{4.2pt}
\begin{tabular}{P{0.08\textwidth} P{0.24\textwidth} P{0.22\textwidth} P{0.35\textwidth}}
\toprule
Topology & Observation geometry & Principal opportunity & Dominant evaluation requirements \\
\midrule
Monostatic & STx and SRx are co-located. Propagation delay corresponds to the round-trip distance to the target. & Tight transmitter--receiver synchronization and direct geometric interpretation & Transmitter leakage and self-interference, near-range masking, dynamic range, target RCS, array pattern, and CPI \\
Bistatic & STx and SRx are separated. Propagation delay corresponds to the total transmitter--target--receiver path length. & Spatial diversity and reduced co-located self-interference & Absolute timing and frequency alignment, node locations, bistatic RCS and Doppler, two-leg path loss, and transmitter--receiver pairing \\
Multistatic & Several STx--SRx pairs observe a common area or target & Coverage extension, geometric diversity, and robust localization or tracking & Cross-node calibration, measurement association, correlated errors, transport latency, fusion rule, missing observations, and inter-node interference \\
\bottomrule
\end{tabular}
\end{table*}

In monostatic sensing, the STx and SRx are colocated at the same radio node. This arrangement simplifies the reference geometry, but the desired echo can be many orders of magnitude weaker than transmitter leakage. Antenna isolation, analog and digital cancellation, clutter suppression, and receiver dynamic range must therefore be included in the assumed operating point~\cite{niu_2025_iotj}. In bistatic sensing, the STx and SRx are located at different radio nodes. The spatial separation can provide useful observation geometry and avoid direct monostatic coupling, but it introduces two-leg delay, bistatic Doppler, transmitter and receiver synchronization requirements, and a target response that depends on both incidence and observation angles. Multistatic sensing uses multiple STx and SRx pairs to observe a common target or area. It can improve coverage and reduce geometric ambiguity, provided that the measurements are correctly associated and that their errors, visibility events, and common biases are not assumed to be independent by default~\cite{li_2024_twc,meng_2025_wc,meng_2025_twc}.

The placement of sensing processing introduces a second architectural
dimension. In a \emph{network-centric} realization, radio nodes transfer
observations or intermediate results to an infrastructure function for fusion,
tracking, and service delivery. In a \emph{device-centric} realization, the UE
performs most sensing processing locally. A \emph{hybrid} realization retains
time-critical processing near the receiver and transfers higher-level
information for network-wide fusion.

The exchanged information can range from in-phase/quadrature samples to
intermediate feature maps and object or track reports. Moving toward
higher-level reports reduces transport requirements but incorporates more
local processing decisions, including detection, association, filtering, and
uncertainty estimation. Section~IV-B describes how the Release~20 study
instantiates this general hierarchy through standardized reporting
abstractions.

\begin{figure*}[!t]
\centering
\includegraphics[width=0.76\textwidth]{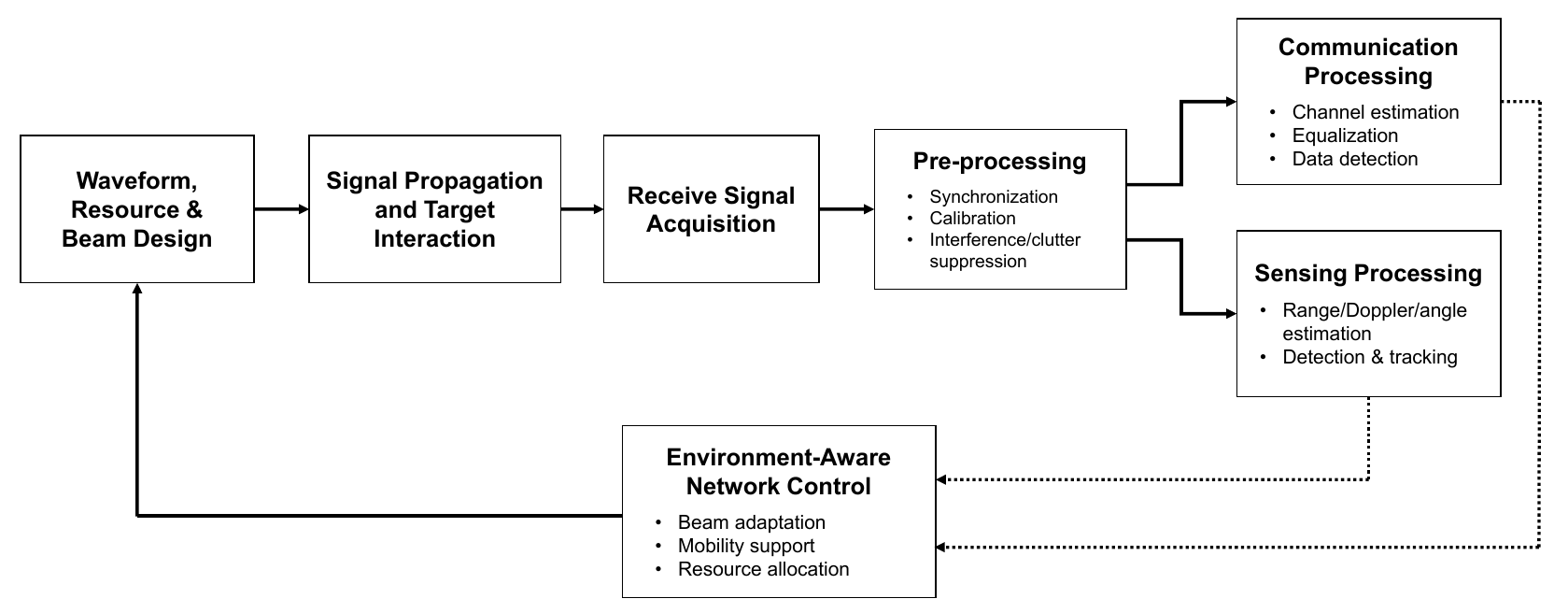}
\caption{Generic cellular ISAC processing loop. A common radio front end supports communication and sensing processing, while link and object information feed environment-aware network control. The feedback path couples sensing quality to subsequent waveform, resource, and beam decisions.}
\label{fig:comst_processing}
\end{figure*}

Fig.~\ref{fig:comst_processing} summarizes the corresponding processing loop. Waveform, resource, and beam design determine the available bandwidth, coherent observation time, aperture, illumination power, and spatial coverage. After propagation and target interaction, acquisition and common preprocessing provide synchronization, calibration, and interference or clutter suppression. The communication branch estimates the link and recovers data, whereas the sensing branch estimates physical parameters before detection, association, localization, classification, or tracking~\cite{zhang_2021_jstsp,shi_2022_jsac}. The resulting information can then support environment-aware control. This feedback path is essential because later beam and resource decisions alter the next observation.

\subsection{Channel, Target, and Interference Abstractions}

At an ISAC sensing receiver, the baseband signal can contain target-dependent propagation, background multipath and clutter, residual self-interference, interference from other radio links, and receiver noise. For one pair of STx and SRx, these contributions can be represented as
\begin{equation}
\resizebox{0.91\columnwidth}{!}{$\displaystyle
\mathbf{y}_{\mathrm{s}}(t)=
\left[
\sum_{k=1}^{K}\mathbf{H}_{\mathrm{tar}}^{(k)}(t,\tau)
+\mathbf{H}_{\mathrm{bg}}(t,\tau)
\right]*\mathbf{x}_{\mathrm{s}}(t)
+\mathbf{z}_{\mathrm{SI}}(t)
+\mathbf{z}_{\mathrm{net}}(t)
+\mathbf{n}(t).
$}
\label{eq:isac_received_signal}
\end{equation}
Here, $*$ denotes convolution over the propagation-delay variable $\tau$, defined by $[\mathbf{H}*\mathbf{x}](t)=\int_{-\infty}^{\infty}\mathbf{H}(t,\tau)\mathbf{x}(t-\tau)\,d\tau$. The vectors $\mathbf{x}_{\mathrm{s}}(t)\in\mathbb{C}^{N_{\mathrm{T}}}$ and $\mathbf{y}_{\mathrm{s}}(t)\in\mathbb{C}^{N_{\mathrm{R}}}$ stack the complex baseband signals associated with the $N_{\mathrm{T}}$ STx and $N_{\mathrm{R}}$ SRx antenna ports, respectively. For each time $t$ and delay $\tau$, the matrices $\mathbf{H}_{\mathrm{tar}}^{(k)}(t,\tau)$ and $\mathbf{H}_{\mathrm{bg}}(t,\tau)$ belong to $\mathbb{C}^{N_{\mathrm{R}}\times N_{\mathrm{T}}}$ and represent the time-varying target-dependent and background MIMO channel impulse responses, respectively, while preserving the spatial and polarimetric responses required for array processing. The final three terms denote residual self-interference, interference from other radio links, and receiver noise, respectively. The target and background channel separation adopted in Clause~7.9 of TR~38.901 is a standards-specific realization of this general abstraction~\cite{ref11}.

For scattering point $p$ associated with target $k$, the direct line-of-sight (LOS) bistatic geometric delay is
\begin{equation}
\tau_{k,p}^{\mathrm{LOS}}(t)=
\frac{
\left\lVert\mathbf{p}_{k,p}(t)-\mathbf{p}_{\mathrm{STx}}(t)\right\rVert+
\left\lVert\mathbf{p}_{k,p}(t)-\mathbf{p}_{\mathrm{SRx}}(t)\right\rVert
}{c},
\label{eq:bistatic_delay}
\end{equation}
where $\mathbf{p}_{k,p}(t)$, $\mathbf{p}_{\mathrm{STx}}(t)$, and
$\mathbf{p}_{\mathrm{SRx}}(t)$ denote the positions of scattering point
$p$ on target $k$, the STx, and the SRx, respectively, and $c$ is the
speed of light. Equation~\eqref{eq:bistatic_delay} relates the propagation
delay to the bistatic range, which is the sum of the STx-to-target and
target-to-SRx distances. For colocated STx and SRx, this relationship
reduces to twice the distance between the radio node and the scattering
point divided by $c$.

The Doppler shift associated with the same propagation path is
\begin{equation}
f_{\mathrm{D},k,p}(t)
=
-f_{\mathrm{c}}\frac{\mathrm{d}\tau_{k,p}^{\mathrm{LOS}}(t)}{\mathrm{d}t}
=
-\frac{1}{\lambda}\frac{\mathrm{d}L_{k,p}(t)}{\mathrm{d}t},
\label{eq:bistatic_doppler}
\end{equation}
where $L_{k,p}(t)=c\tau_{k,p}^{\mathrm{LOS}}(t)$ is the total bistatic
path length, $f_{\mathrm{c}}$ is the carrier frequency, and
$\lambda=c/f_{\mathrm{c}}$ is the wavelength. The adopted sign convention
assigns positive Doppler to a decreasing total path length. The bistatic
Doppler represents the sum of the relative-velocity projections along
the two propagation legs. For monostatic sensing, it reduces to twice
the radial-velocity projection divided by the wavelength.

Additional propagation paths may include relative delay and excess delay
on either leg. Localization and velocity evaluation therefore require
absolute geometry, target-dependent delay and Doppler, and spatially
consistent channel evolution~\cite{wei2025isacchannel}.

The coefficient of a target path is governed jointly by propagation loss, antenna response, polarization, and RCS. RCS quantifies the effective electromagnetic scattering strength of a target~\cite{10494709}. For monostatic observation it characterizes the backscattered field toward the co-located transmitter and receiver, whereas bistatic RCS depends separately on the incident and observation directions~\cite{wei2025isacchannel,10789723}. It can be represented by a deterministic angular signature when target orientation is known or by a target-conditioned random process when the detailed state is unavailable~\cite{wei2025isacchannel,4102863}. In either form, RCS is not a fixed gain: it may vary with frequency, target class, aspect angle, bistatic angle, polarization, and time, thereby changing detection and sensing reliability~\cite{10548468}.

UAVs are important ISAC targets in inspection, delivery, emergency response, and aerial-connectivity scenarios~\cite{liu_2023_spm}, but their scattering is particularly difficult to reduce to one scalar. The fuselage, arms, rotors, landing gear, and payload housing form multiple scattering components whose relative phases change with aspect and frequency. A UAV may consequently be strongly visible in one direction and weak after a small change in attitude or sensing geometry.

Array-based evaluation requires element positions, radiation patterns,
and beamforming weights to represent spatial processing. A fixed scalar
RCS can be combined with these array responses to evaluate array gain,
beam scanning, and angle estimation for an idealized point target.
Evaluating polarization-dependent performance additionally requires
a polarimetric target-response model, while realistic UAV scattering
may require responses that vary with frequency, orientation, and
observation geometry. The measurements in Section~\ref{sec:testbeds} illustrate these target-dependent variations.

Target and background channel contributions should be distinguished during performance evaluation, even when both are generated using a common geometry-based model. Target paths are associated with an object whose state is to be inferred, whereas background paths may produce clutter peaks, masking, or multipath-induced ghost objects. Spatially or temporally correlated clutter cannot be replaced by independent white noise without changing the detector operating characteristic. Likewise, target visibility across several TRPs can be correlated because the links share blockage, target orientation, or large-scale environmental conditions.

Interference in arises from coupling within a sensing node and
transmissions from other radio links. Residual self-interference can limit
simultaneous monostatic sensing, depending on isolation and cancellation
performance. Mutual sensing interference arises when neighboring sensing nodes use overlapping resources, while communication--sensing cross-interference occurs when the two functions share resources without sufficient isolation. External transmissions may provide additional sensing illumination when waveform knowledge, synchronization, and calibration permit. Otherwise, their effects should be modeled explicitly or through a validated equivalent-noise approximation~\cite{niu_2025_iotj}.

\subsection{Evaluation Levels, Metrics, and Statistical Protocols}
\label{subsec:evaluation_metrics}

No single evaluation approach can capture every aspect of ISAC operation and performance. Table~\ref{tab:isac_evaluation_levels} summarizes three complementary evaluation approaches. Link-level simulation resolves waveforms, arrays, channels, detection, and estimation. System-level simulation abstracts those operations sufficiently to simulate many nodes, drops, scheduling intervals, and mobility events. End-to-end implementation evaluation determines whether measurement transport, processing delay, interfaces, and control logic preserve the expected benefit. Results should be interpreted within the scope actually evaluated.

\begin{table}[!t]
\caption{ISAC evaluation approaches and their scope}
\label{tab:isac_evaluation_levels}
\centering
\footnotesize
\setlength{\tabcolsep}{3.2pt}
\begin{tabular}{P{0.15\columnwidth} P{0.33\columnwidth} P{0.42\columnwidth}}
\toprule
Evaluation approach & Modeled components & Applicable performance metrics \\
\midrule
Link-level simulation & Waveform, array, channel, detector, estimator & Detection, false alarm, resolution, and conditional estimation error \\
System-level simulation & Layout, mobility, interference, scheduling, association, abstraction of link behavior & Coverage, user distributions, handover, resource use, and network-level reliability \\
End-to-end implementation & Interfaces, processing and transport delay, protocol state, hardware impairments & Feasibility, latency, control-loop behavior, and implementation constraints \\
\bottomrule
\end{tabular}
\end{table}

Detection probability and false-alarm probability (FAP) jointly characterize whether a sensing system detects present targets while avoiding reports unsupported by the reference target configuration. Detection probability is evaluated for targets that are present, whereas false-alarm probability characterizes erroneous detections. The detection threshold, evaluated population, association rule, and counting method must be specified because different definitions can produce different FAPs.

Evaluation of position and velocity estimation accuracy should specify how missed targets, unassociated detections, and tracking failures are treated. An error cumulative distribution function (CDF) computed only for correctly detected and associated targets represents \emph{conditional accuracy} and excludes missed detections. Detection probability and conditional estimation error should therefore be reported together. Mean and root-mean-square errors are useful for calibration, whereas the median and tail percentiles reveal non-Gaussian error behavior. Tracking evaluation additionally requires track continuity, initiation and deletion rules, data association, and potentially object-set metrics. 

When sensing information is used to assist communication, the resulting communication performance must be evaluated using metrics defined independently of sensing accuracy. User-throughput evaluation should specify the modulation and coding scheme (MCS), which determines the modulation order and coding rate, together with the MCS selection rule, link-to-system mapping, treatment of retransmissions and outages, allocated resource share, and averaging interval. For sensing-assisted mobility evaluation, handover failure should be defined using an explicit radio-link or protocol criterion, while handover interruption time should have clearly stated start and end events. A sensing-assisted policy should be compared with a conventional policy under matched radio resources and equivalent control conditions. Otherwise, an apparent improvement may result from a more permissive handover trigger rather than from better environmental information.

Reliable comparison requires a predefined statistical evaluation procedure. Calibration and evaluation samples should be separated, and competing methods should be assessed using matched operating points and paired realizations whenever possible. Sample counts and statistical uncertainty should be reported, particularly for rare events such as false alarms and handover failures.

\subsection{Communication--Sensing Tradeoffs}

ISAC creates coupled tradeoffs across time, frequency, spatial, and power
resources, the locations and computational complexity of processing functions, and the capacity, latency, and granularity of sensing-data exchange~\cite{liu_2022_cst,wei_2023_iotj}.
Increasing the signal bandwidth provides finer delay resolution and consequently improves range resolution. Extending the
CPI improves Doppler resolution. Enlarging
the effective antenna aperture improves angular resolution. These choices
can reduce the resources available for communication, increase observation
latency, restrict spatial coverage, and impose tighter phase-coherence
requirements. Higher illumination power can improve the sensing
signal-to-noise ratio (SNR) but may also increase self-interference and
interference to neighboring links. Detailed and frequent sensing-data
exchange can support multi-node fusion and maintain information freshness,
but it increases fronthaul, backhaul, and core-network load as well as
transport and processing latency. Compact or less frequent reports reduce
this overhead but may decrease information fidelity and increase the age of sensing information.

The design objective is therefore to balance sensing quality with communication performance rather than optimize sensing accuracy or throughput independently. An effective design seeks a feasible operating region subject to communication quality-of-service, sensing reliability, latency, energy consumption, and implementation constraints. Resource-allocation studies formalize several aspects of this multi-objective problem~\cite{zong_2025_tvt,yang_2024_twc}. Network-level evaluation must additionally represent the feedback loop between sensing observations and subsequent radio decisions. For sensing-assisted communication, the net benefit can be viewed as the communication loss avoided through sensing assistance minus the radio-resource, processing, reporting, and decision costs required to acquire and use the sensing information. An ideal sensing-assisted method consequently provides an upper performance reference, whereas a practical method must account for observation errors and information delay.

%% file: sections/04_standardization.tex
\section{ISAC Standardization in 3GPP}
\label{sec:standardization}

The 3GPP standardization of cellular ISAC has progressed through coordinated work across multiple technical and architectural layers rather than through a single specification. Release~19 defined service requirements and a target-aware sensing channel modeling foundation. Release~20 then defined an NR sensing evaluation baseline, studied RAN-side sensing control and information transfer, and introduced a service architecture for sensing management and application access to sensing information. These activities have different status: a study report defines assumptions and evaluates feasibility, whereas a normative technical specification defines interoperable behavior. We preserve this distinction while connecting the service, radio, RAN, and core-network perspectives.

\subsection{Release~19 Foundations}

Release~19 established the service and channel-modeling foundations for cellular ISAC. These activities identified sensing use cases and requirements, defined representative deployment scenarios and sensing modes, and introduced target-aware channel models.

\subsubsection{Services and Use Cases}

The Release~19 SA1 study documented in TR~22.837 examined how the 5G system could support sensing services using 3GPP radio signals and infrastructure~\cite{tr_22_837}. Building on this study, TS~22.137 specifies the Release~19 functional and performance requirements for 5G wireless sensing services~\cite{ts22137}. The objective was not limited to estimating a target range or velocity. It encompassed the complete process of requesting a sensing service for a given geographical area and time interval, choosing authorized sensing entities, collecting sensing data, producing a result that meets defined performance criteria, and making that result available to an application in accordance with operator policy.

\begin{figure*}[!t]
\centering
\includegraphics[width=0.66\textwidth]{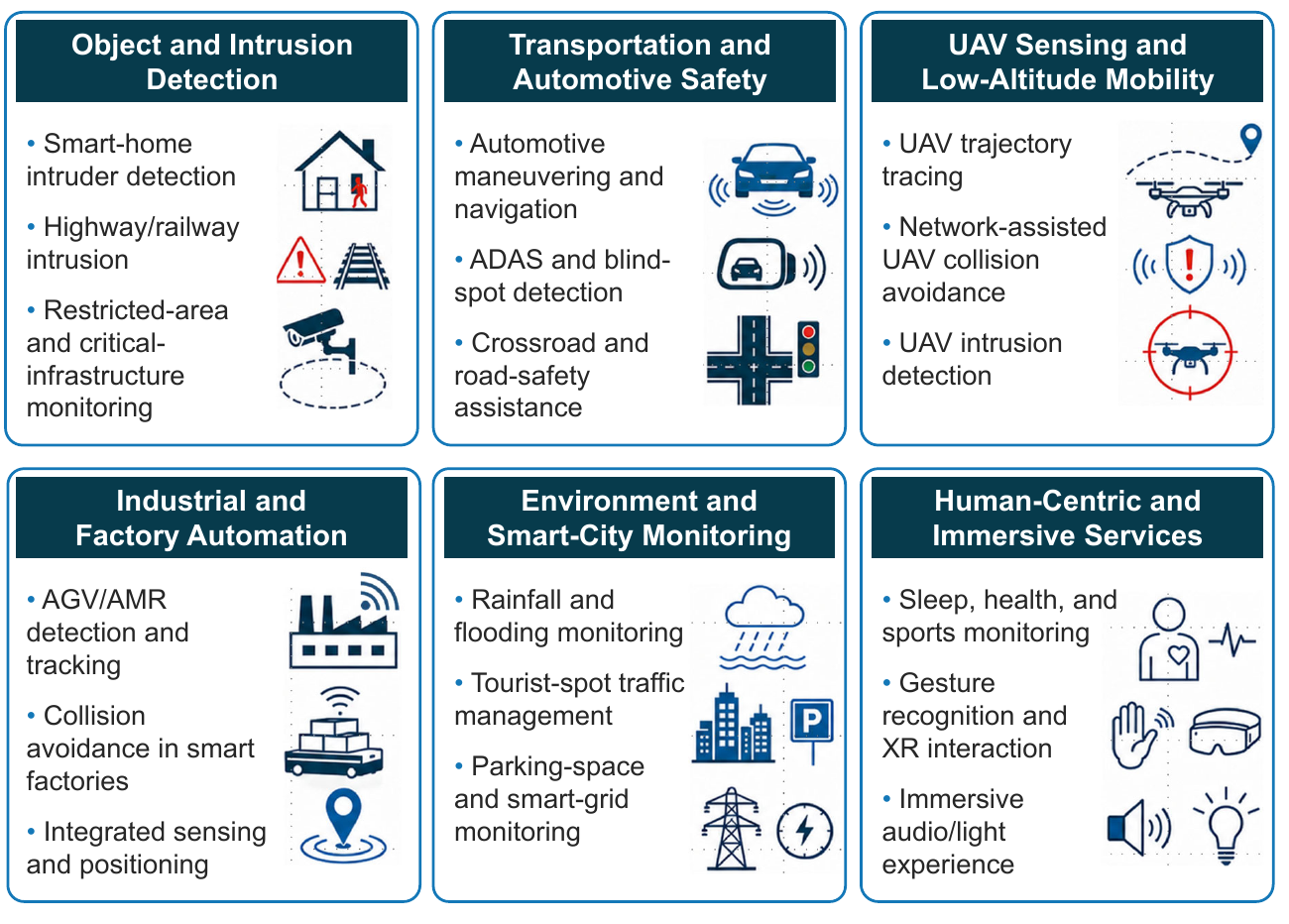}
\caption{Representative ISAC service scenarios identified in 3GPP TR~22.837.}
\label{fig:comst_services}
\end{figure*}

Fig.~\ref{fig:comst_services} categorizes representative applications into six domains. Object and intrusion detection covers monitoring in smart homes, highways, railways, and restricted areas. Transportation-oriented services support vehicle maneuvering, blind-spot detection, road safety, and navigation, while UAV sensing enables trajectory tracing, collision avoidance, and restricted-airspace monitoring. Industrial applications address automated guided vehicle (AGV) and autonomous mobile robot detection, worker protection, and integrated sensing and positioning. Environmental and smart-city services encompass rainfall, flooding, parking, traffic, and infrastructure monitoring. Human-centric applications extend to contactless health and activity monitoring, gesture recognition, and immersive interaction.

Across these applications, the recurring requirements are service authorization, target-area description, sensing activation and termination, result quality, reporting periodicity, service continuity, prioritization, and exposure. The study also recognizes that sensing results may reveal information about people, vehicles, private areas, or critical infrastructure. Authorization, privacy, consent, integrity, and charging are therefore service requirements rather than optional implementation details. It further allows a sensing service to combine 3GPP observations with non-3GPP information such as camera, LiDAR, or dedicated-radar data. This broad service vision is important because the initial radio and architecture baselines support only a restricted subset of the use cases identified by the Release~19 SA1 study.

\subsubsection{Deployment Scenarios and Sensing Modes}

The Release~19 RAN1 study translated the service-level use cases into representative targets, environments, and sensing geometries for channel modeling. Table~\ref{tab:comst_release19_scenarios} summarizes five principal deployment scenario groups: UAV, human, automotive, industrial AGV, and hazardous-object sensing. They span aerial and ground targets, indoor and outdoor propagation, static and mobile objects, and target responses with different sizes, materials, motion signatures, and safety implications~\cite{ref11}. The deployment environments include urban microcell (UMi), urban macrocell (UMa), and rural macrocell (RMa) scenarios.

\begin{table*}[!t]
\caption{Representative Release~19 ISAC deployment scenarios}
\label{tab:comst_release19_scenarios}
\centering
\footnotesize
\setlength{\tabcolsep}{4pt}
\begin{tabular}{P{0.15\textwidth} P{0.40\textwidth} P{0.40\textwidth}}
\toprule
Scenario & Representative environment and targets & Principal sensing objectives \\
\midrule
UAV sensing & Urban, suburban, and rural macro deployments; low-altitude aerial targets & Detection, trajectory tracking, collision avoidance, and restricted-airspace monitoring \\
Human sensing & Indoor office, residential, factory, and outdoor public areas & Presence, activity, gesture, and contactless health or safety monitoring \\
Automotive sensing & UMi, UMa, and RMa roads and intersections; vehicles and vulnerable road users & Maneuvering, navigation, blind-spot detection, and crossroad safety \\
AGV and factory sensing & Factories and warehouses; AGVs, robots, workers, and infrastructure & Detection, tracking, collision avoidance, and integrated sensing-positioning support \\
Hazardous-object sensing & Roads and railways; debris, animals, pedestrians, fallen objects, or unauthorized vehicles & Hazard detection, localization, and infrastructure protection \\
\bottomrule
\end{tabular}
\end{table*}

\begin{figure}[!t]
\centering
\includegraphics[width=\columnwidth]{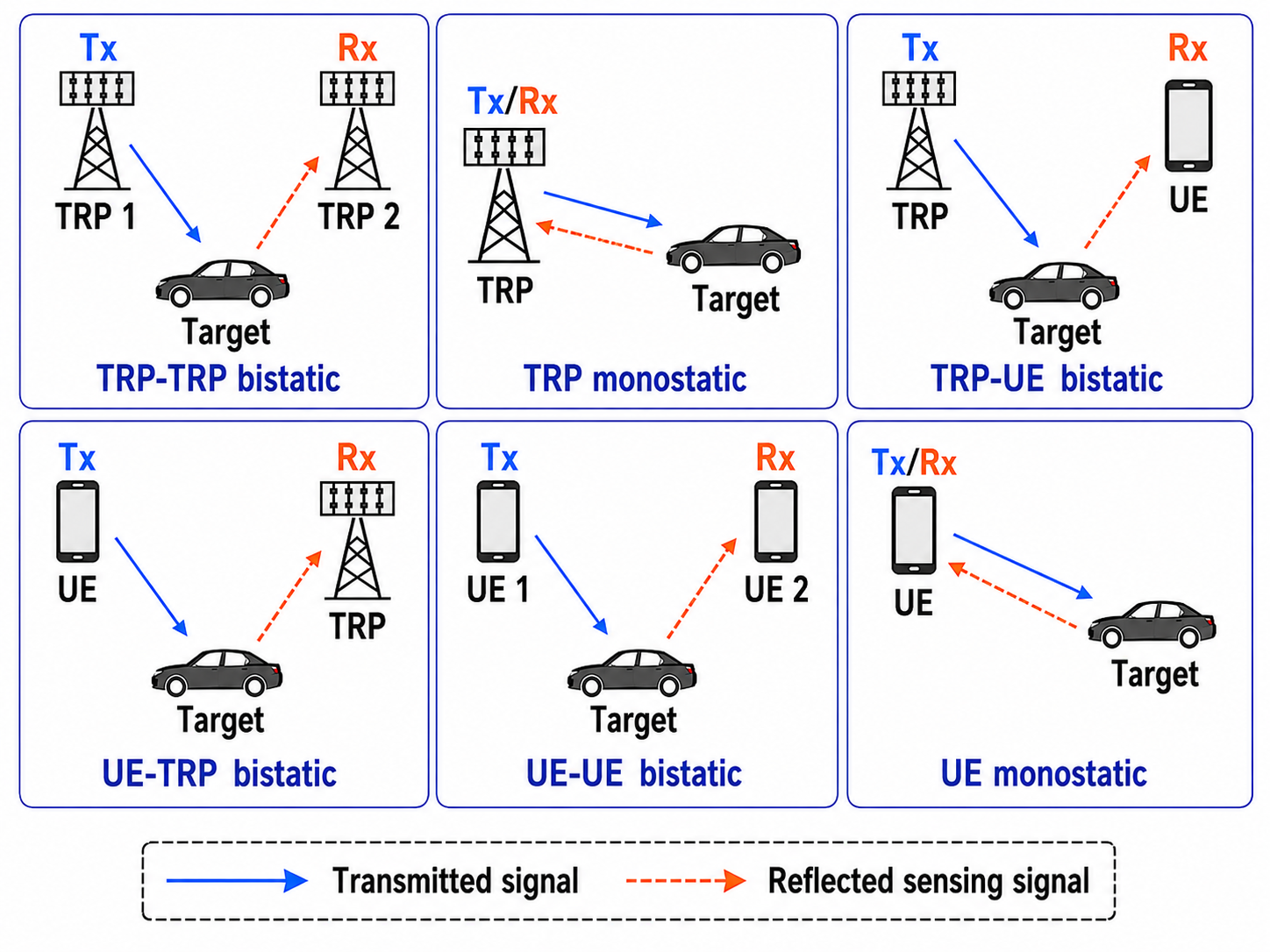}
\caption{TRP- and UE-based sensing modes considered in the 3GPP ISAC channel-modeling study.}
\label{fig:comst_sensing_modes}
\end{figure}

The sensing mode is determined by whether the sensing transmitter and receiver are TRPs or UEs and whether they are co-located or not. Fig.~\ref{fig:comst_sensing_modes} shows six modes: TRP--TRP bistatic, TRP monostatic, TRP--UE bistatic, UE--TRP bistatic, UE--UE bistatic, and UE monostatic sensing. These modes differ in their sensing geometries and in their requirements for self-interference suppression, inter-node synchronization, UE participation, and network control. TRP monostatic sensing offers infrastructure control and a direct round-trip geometry but requires strong self-interference suppression. TRP--TRP bistatic sensing separates the transmitter and receiver, thereby avoiding monostatic self-interference at the sensing receiver, but it requires inter-node synchronization and two-leg target-channel modeling~\cite{willis2005bistatic,ref11}. UE-involved modes can extend sensing coverage and provide additional observation geometries near the target. Their practical realization must also account for UE capability and energy limitations, authorization, mobility, and sensing-reporting procedures~\cite{tr_22_837,gonzalez_2024_pieee}.

These deployment scenarios and sensing modes define the physical conditions that an ISAC channel model must represent. The target class and surrounding environment determine the target response, background scattering, and clutter characteristics. The monostatic or bistatic configuration determines the transmitter--target--receiver geometry and the corresponding delays, angles, Doppler shifts, and propagation losses. UE-involved sensing additionally introduces mobile endpoints and time-varying observation geometries. Conventional communication channel models do not explicitly preserve all these target-dependent properties, which motivates the ISAC channel-model extensions.

\subsubsection{ISAC Channel Modeling}

The principal Release~19 RAN1 radio contribution is the ISAC extension now reflected in Clause~7.9 of TR~38.901~\cite{ref11}. The legacy geometry-based stochastic model represents a communication link through statistically generated clusters, rays, delays, powers, angles, polarization, blockage, and mobility. Sensing requires additional structure because the receiver must attribute part of the received signal to a physical target and preserve the absolute geometry used to estimate its state.

The model therefore distinguishes a \emph{target channel} from a \emph{background channel}. A sensing target is represented by one or more scattering points of a sensing target (SPSTs). Each SPST produces target-related paths with a location, velocity, RCS, polarization response, absolute delay, angles, and Doppler evolution. Environment objects and conventional multipath generate the background contribution, including clutter and non-target reflections. The complete received response combines these contributions, while their separate generation enables target-related and background-induced effects to be distinguished during performance evaluation.

\begin{figure}[!t]
\centering
\includegraphics[width=\columnwidth]{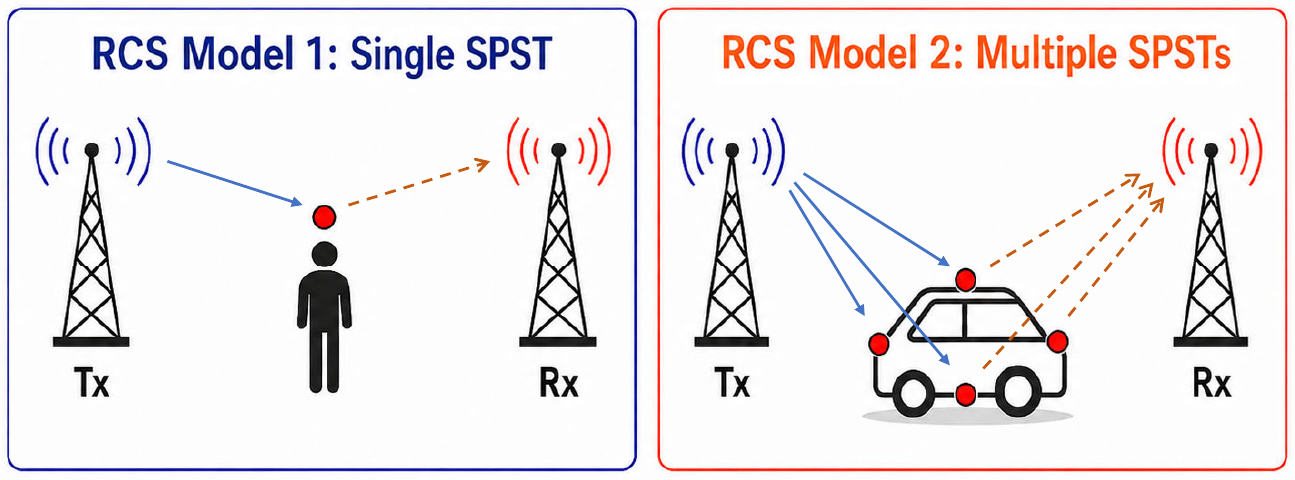}
\caption{Single- and multiple-SPST RCS representations used in ISAC channel modeling.}
\label{fig:comst_rcs_model}
\end{figure}

RCS provides the target-dependent power scaling. A compact target can be approximated by a single SPST with an assigned
RCS, whereas an extended target can be represented by multiple SPSTs.
In either representation, the assigned RCS may depend on frequency,
orientation, observation direction, and polarization. The latter preserves changes in target visibility with incidence angle, departure angle, carrier frequency, polarization, and target orientation. This distinction is essential for UAVs, vehicles, and AGVs whose frames, rotors, body panels, and payloads do not scatter uniformly. RCS is combined with two-leg propagation loss, transmit and receive antenna responses, the polarimetric coupling matrix, array phases, and target-induced Doppler. With explicit antenna-array responses, a fixed scalar RCS can support
evaluation of panel-size effects for an idealized point target.
Direction- and polarization-dependent target responses are additionally
needed to assess how these effects vary with target orientation,
observation geometry, and polarization configuration. Fig.~\ref{fig:comst_rcs_model} compares the single- and multiple-SPST representations.

\begin{figure}[!t]
\centering
\includegraphics[width=\columnwidth]{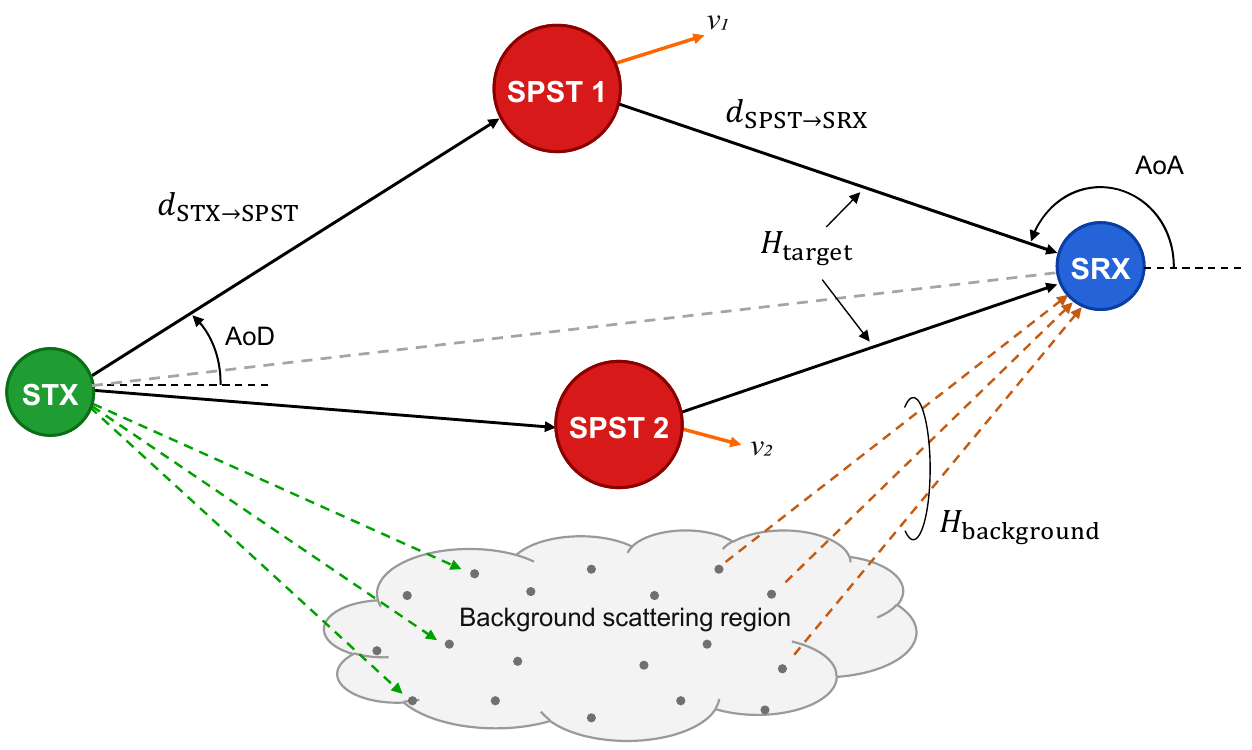}
\caption{Conceptual bistatic sensing geometry with target-related SPST paths and a background-scattering contribution.}
\label{fig:comst_geometry}
\end{figure}

Building on the direct LOS delay through an SPST introduced in~\eqref{eq:bistatic_delay}, general target paths incorporate relative cluster or ray delays and non-line-of-sight (NLOS) excess delays along the transmitter-to-SPST and SPST-to-receiver propagation segments. Their time evolution accounts for the motion of the transmitter, receiver, target, and individual scattering points, thereby supporting target macro-motion and parameterized micro-motion such as rotor or wheel dynamics. Spatial consistency ensures that delays, angles, powers, visibility, and RCS evolve continuously. Independently generated channel snapshots cannot provide this continuity and are therefore unsuitable for tracking or sensing-assisted mobility evaluation. Fig.~\ref{fig:comst_geometry} illustrates the corresponding target and background geometries.

\begin{figure}[!t]
\centering
\begingroup
\makeatletter
\ifdefined\Gread@transgrouptrue
  \let\Gread@transgrouptrue\Gread@transgroupfalse
\fi
\makeatother
\includegraphics[width=\columnwidth]{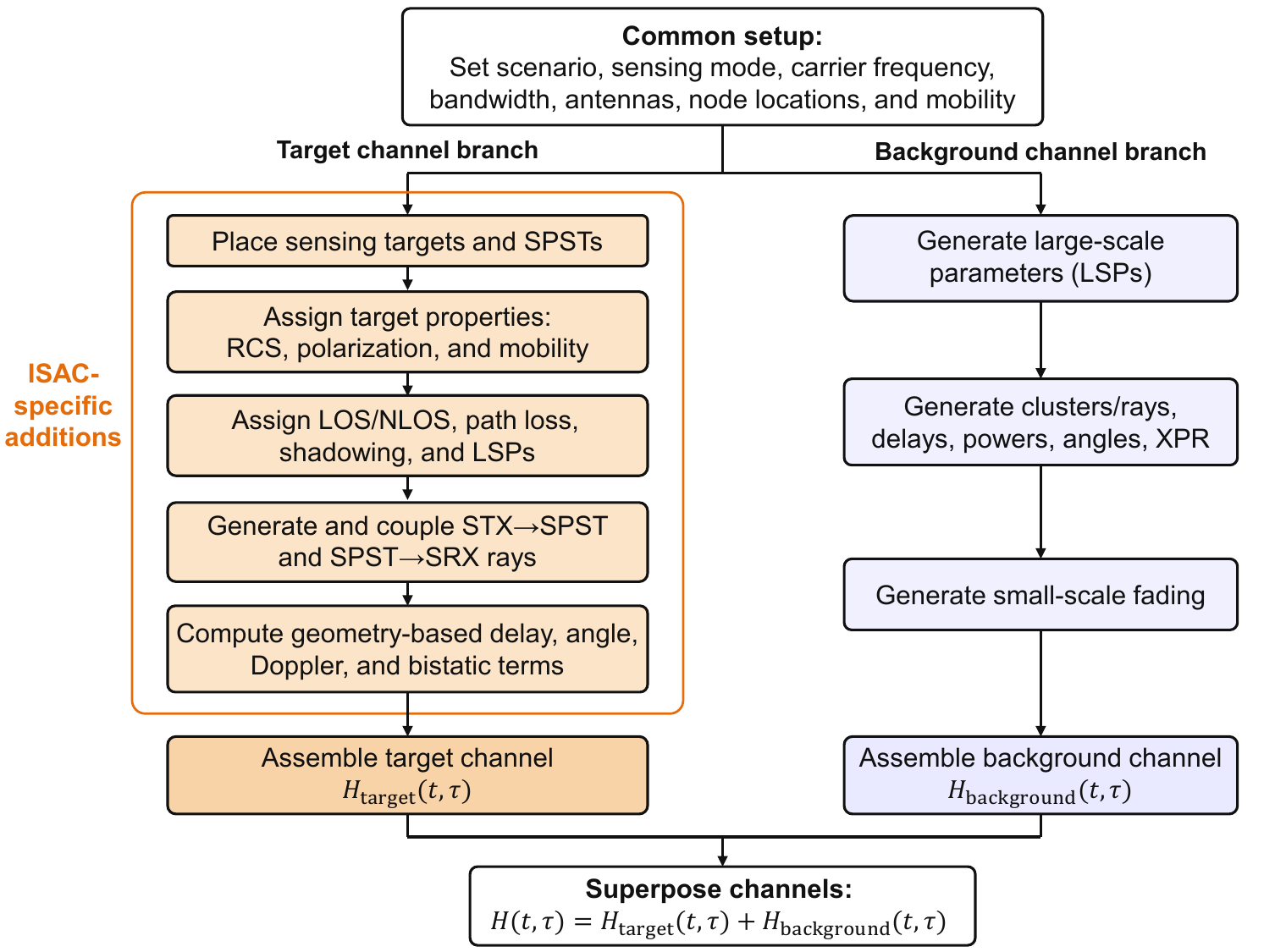}
\endgroup
\caption{Target- and background-channel generation flow for the Clause~7.9 ISAC channel model.}
\label{fig:comst_channel_generation}
\end{figure}

Fig.~\ref{fig:comst_channel_generation} summarizes the generation flow. After common scenario, node, array, and mobility initialization, the background branch reuses the applicable TR~38.901 communication-channel procedure. The target branch places the sensing targets and SPSTs, assigns RCS and mobility, generates the transmitter-to-SPST and SPST-to-receiver legs, couples the legs, and computes geometry-consistent delay, angle, Doppler, phase, and polarization. The target and background coefficients are then superposed at the receiver. This reuse preserves compatibility with established 3GPP propagation scenarios while adding the object-centric quantities required by sensing.

TR~38.901 Clause~7.9 defines a channel-generation framework rather than an end-to-end sensing system. Detector-threshold selection, multi-target peak association, multi-TRP fusion, and reporting-interface design fall outside its scope. Consequently, a 3GPP-aligned evaluation requires more than the adoption of standardized scenario labels. A reproducible study must additionally specify the RCS model, target distribution, antenna arrays, polarization, background cancellation, self-interference, receiver processing, and statistical protocol.

\subsection{Release~20 NR Sensing Study}

Release~20 extended the Release~19 target-aware channel-modeling foundation into coordinated studies of radio feasibility, RAN procedures, and core-network support for sensing services. RAN1 established in TR~38.765 an initial evaluation framework for gNB-based UAV sensing, including radio assumptions, sensing metrics, and performance objectives. RAN3 examined sensing-task control and information transfer between the gNB and SenF. In parallel, SA2 studied the core-network functions and procedures required to manage sensing requests and deliver results to authorized applications. Together, these activities provide an initial path from radio-level sensing to network-supported services, although their scope remains narrower than the Release~19 service vision.

\subsubsection{Evaluation Scope and Assumptions}

TR~38.765 includes the evaluation of gNB-centric NR sensing with UAV detection and tracking as the initial baseline use case~\cite{tr38765}. As specified in Clause~6.1 of TR~38.765, the evaluation procedure configures the scenario, drops targets, selects sensing transmitters and receivers, generates the target and background channels, propagates a sensing signal, processes each receiver observation, optionally fuses measurements or tracks, and computes target-level metrics. This order connects the Clause~7.9 channel model to a reproducible detection and estimation evaluation workflow.

Table~\ref{tab:comst_tr38765_baseline} summarizes the principal assumptions and objectives. The baseline uses the UMa scenario with aerial vehicles (UMa-AV), with 500~m inter-site distance, 4 or 4.9~GHz carrier frequency, simultaneous sensing transmission and reception, dual-polarized base station (BS) panels, five targets per sector at the center site, and UAV heights from 25 to 300~m. Baseline configurations~1 and~2 mainly differ in maximum BS transmit power, 52 and 37~dBm, respectively. 

\begin{table}[!t]
\caption{Release~20 UAV sensing evaluation baseline and objectives}
\label{tab:comst_tr38765_baseline}
\centering
\footnotesize
\setlength{\tabcolsep}{3pt}
\begin{tabular}{P{0.25\columnwidth} P{0.7\columnwidth}}
\toprule
Item & Baseline assumption or objective \\
\midrule
Scenario & UMa-AV, 500~m inter-site distance \\
Operation & Sensing Tx/Rx operating simultaneously \\
Carrier frequency & 4 or 4.9~GHz \\
Maximum BS power & 52~dBm (configuration~1) or 37~dBm (configuration~2) \\
BS panel & Dual polarized; Tx and Rx $(M,N,P,M_g,N_g,M_p,N_p)=(8,8,2,1,1,4,8)$; $(d_H,d_V)=(0.5,0.8)\lambda$; $+45^\circ/-45^\circ$ polarization \\
Targets & Five per sector at the center site; 25--300~m vertical distribution \\
Detection objectives & Missed detection, Type-1 FAP, and Type-2 FAP each $\leq 5\%$ \\
Estimation objectives & Horizontal and vertical position error $\leq 10$~m and velocity error $\leq 5$~m/s, each at the 90th percentile \\
\bottomrule
\end{tabular}
\end{table}

TR~38.765 divides false alarms into two mandatory metrics~\cite{tr38765}. Type-1 FAP applies to target-absent drops and is the fraction of such drops in which at least one object is detected. Type-2 FAP applies to target-present drops and averages the fraction of detected objects that cannot be associated with any true target, considering drops with at least one detected object. The two FAPs use different conditioning populations and denominators and must therefore be interpreted separately. For target-present drops, the same one-to-one association rule determines missed targets, Type-2 false alarms, and the estimates included in the position and velocity accuracy evaluation.

The objectives in Table~\ref{tab:comst_tr38765_baseline} provide a common feasibility reference. They are not service-level guarantees, because end-to-end performance also depends on scheduling, reporting, processing and transport latency, fusion, tracking, and service exposure. Similarly, a method meeting the target-level objectives in one simulated operating point does not establish coverage across other target classes, bands, geometries, interference conditions, or implementation impairments.

\subsubsection{Network Architecture and Interface}
\label{subsubsec:ran_architecture}

The RAN-side ISAC study builds on the SenF introduced in the SA2 sensing-service architecture and defines the gNB-side interface and procedures for sensing control and information transfer~\cite{tr2370014,tr38765}. Through this interface, a gNB receives requests to initiate, modify, or terminate a sensing task and returns sensing information at the requested reporting level. The SenF can select one or more gNBs according to the target area, sensing capabilities, coverage, and resource availability, while each selected gNB retains control of its local radio resources and performs the required time-critical radio and signal processing.

\begin{figure}[!t]
\centering
\includegraphics[width=0.93\columnwidth]{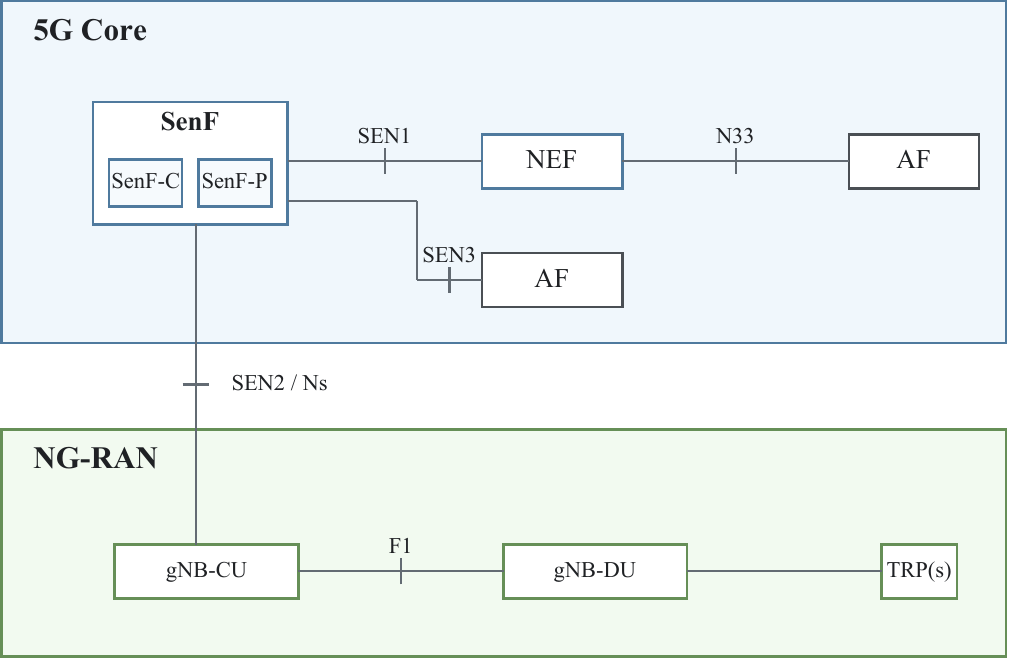}
\caption{Combined RAN and service-architecture view for NR ISAC.}
\label{fig:comst_architecture}
\end{figure}

Fig.~\ref{fig:comst_architecture} combines the RAN and service-architecture
views. On the RAN side, the dedicated logical interface between the SenF and the gNB central unit (gNB-CU) is denoted Ns and carries the Ns Application Protocol (NsAP). In the SA2 reference-point model, the same SenF--gNB relationship is denoted by SEN2. F1 connects the gNB-CU and the gNB distributed unit (gNB-DU). The illustrated gNB-DU--TRP association is left unlabeled because it is an implementation relationship rather than a separately standardized 3GPP interface. The 5G Core (5GC) functions and sensing-service reference points shown in the figure are discussed in the following subsubsection.

For the gNB-based sensing mode, in which no UE participates as a sensing transmitter or receiver, the direct SenF--gNB interface enables sensing control and information transfer without involving the Access and Mobility Management Function (AMF). To support these exchanges over Ns, the study considered Stream Control Transmission Protocol (SCTP)-based, GPRS Tunnelling Protocol User Plane (GTP-U)/User Datagram Protocol (UDP)-based, and WebSocket-based alternatives for transporting control signaling and sensing information over the Ns interface. SCTP provides reliable and ordered delivery while reusing an established
3GPP control-plane transport pattern~\cite{rfc9260,3gpp_ts_38412}. GTP-U over UDP can support efficient transfer of sensing information, while requiring additional handling of reliability, segmentation, congestion control, and packet reordering~\cite{tr38765}. WebSocket provides Transmission Control Protocol (TCP)-based reliability and application-layer flexibility. Based on this evaluation, TR~38.765 concludes that the SCTP-based protocol stack should be adopted for sensing-data transmission in a potential future normative phase. It also presents NsAP over SCTP/IP as the protocol stack for sensing signaling~\cite{tr38765}.

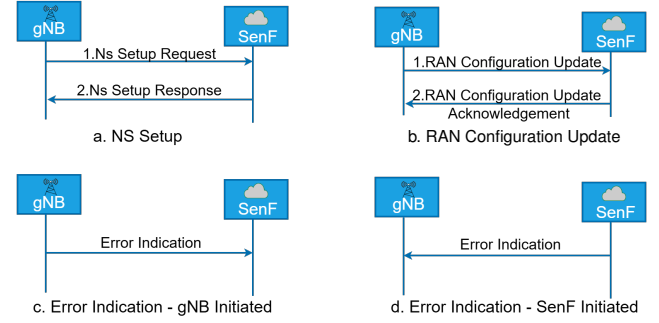
\begin{figure}[!t]
\centering
\input{figures/ns_interface_corrected.tex}
\caption{Ns interface management procedures between the gNB and SenF.}
\label{fig:comst_ns_interface_management}
\end{figure}

NsAP groups its procedures into interface management and sensing information transfer~\cite{tr38765}. Interface management procedures establish and maintain the operational relationship between the gNB and the SenF over the Ns interface. This category comprises Ns Setup, RAN Configuration Update, and Error Indication, as illustrated in Fig.~\ref{fig:comst_ns_interface_management}.

\begin{itemize}
\item \textbf{Ns Setup:} The gNB initiates Ns Setup to establish an operational association with the SenF before sensing tasks are configured. The request provides information about the gNB's supported sensing area and relevant sensing capabilities, enabling the SenF to determine whether the specified target sensing area is covered and whether the gNB can satisfy the requirements of the requested sensing task. The SenF returns an Ns Setup Response when the association is accepted or an Ns Setup Failure otherwise.

\item \textbf{RAN Configuration Update:} The gNB initiates this procedure to convey changes in its sensing-related configuration, including its supported sensing area or capabilities. The SenF returns a RAN Configuration Update Acknowledge when the changes are applied or a RAN Configuration Update Failure when the update cannot be processed.

\item \textbf{Error Indication:} Either the gNB or the SenF initiates this procedure after detecting a protocol error for which no procedure-specific failure response is defined. The Error Indication informs the peer node of the detected error condition.
\end{itemize}

\begin{figure}[!t]
\centering
\includegraphics[width=\columnwidth]{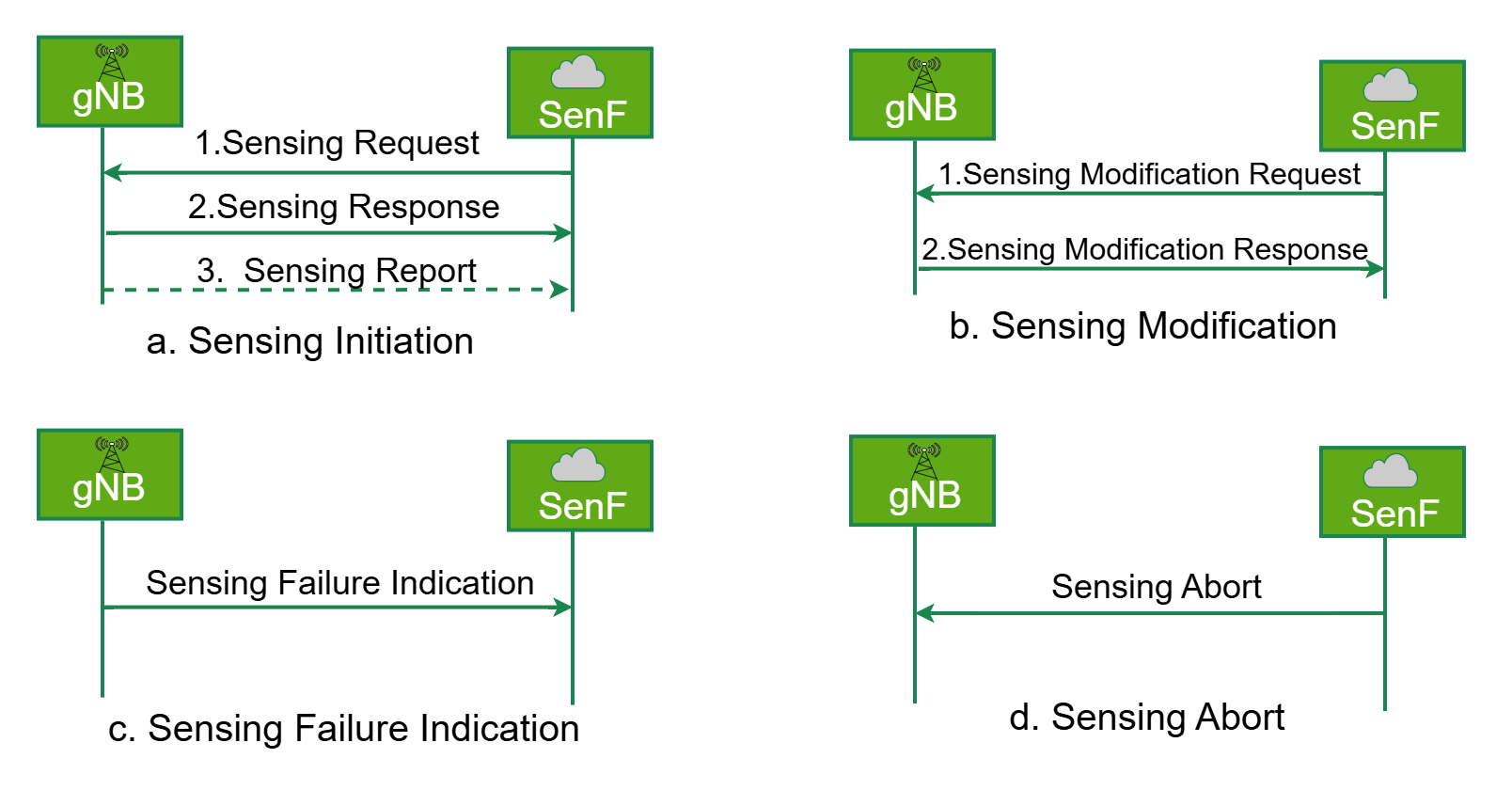}
\caption{Ns sensing-information transfer procedures between the SenF and gNB.}
\label{fig:comst_ns_procedures}
\end{figure}

Sensing information transfer procedures manage the lifecycle of a sensing measurement between the SenF and the gNB. This category comprises Sensing Initiation, Sensing Modification, Sensing Abort, and Sensing Failure Indication, as illustrated in Fig.~\ref{fig:comst_ns_procedures}.

\begin{itemize}
\item \textbf{Sensing Initiation:} The SenF initiates this procedure by sending a Sensing Request that asks the gNB to perform a sensing measurement. The request identifies the sensing measurement and may specify the three-dimensional target sensing area, reporting mode, event conditions, performance requirements, and transport information. The reporting mode can request one-time, periodic, or event-triggered delivery. The gNB returns a Sensing Response when it accepts the task or a Sensing Failure when the request cannot be supported. An accepted task subsequently produces sensing-information reports according to the configured reporting mode.

\item \textbf{Sensing Modification:} The SenF initiates this procedure to update an ongoing sensing measurement. The request identifies the measurement and specifies changes such as the target sensing area, reporting configuration, or performance requirements. The gNB returns a Sensing Modification Response when the changes are accepted or a Sensing Modification Failure when they cannot be applied.

\item \textbf{Sensing Abort:} The SenF initiates this procedure to terminate an ongoing sensing measurement when the task is no longer required or its associated resources need to be released. After receiving the request, the gNB stops the identified measurement and releases the corresponding sensing resources.

\item \textbf{Sensing Failure Indication:} The gNB initiates this procedure when it can no longer maintain an accepted sensing measurement because of resource limitations, configuration changes, or an inability to meet the requested requirements. The indication allows the SenF to modify the task, terminate it, or select another suitable gNB.
\end{itemize}

\begin{figure*}[!t]
\centering
\includegraphics[width=0.95\textwidth]{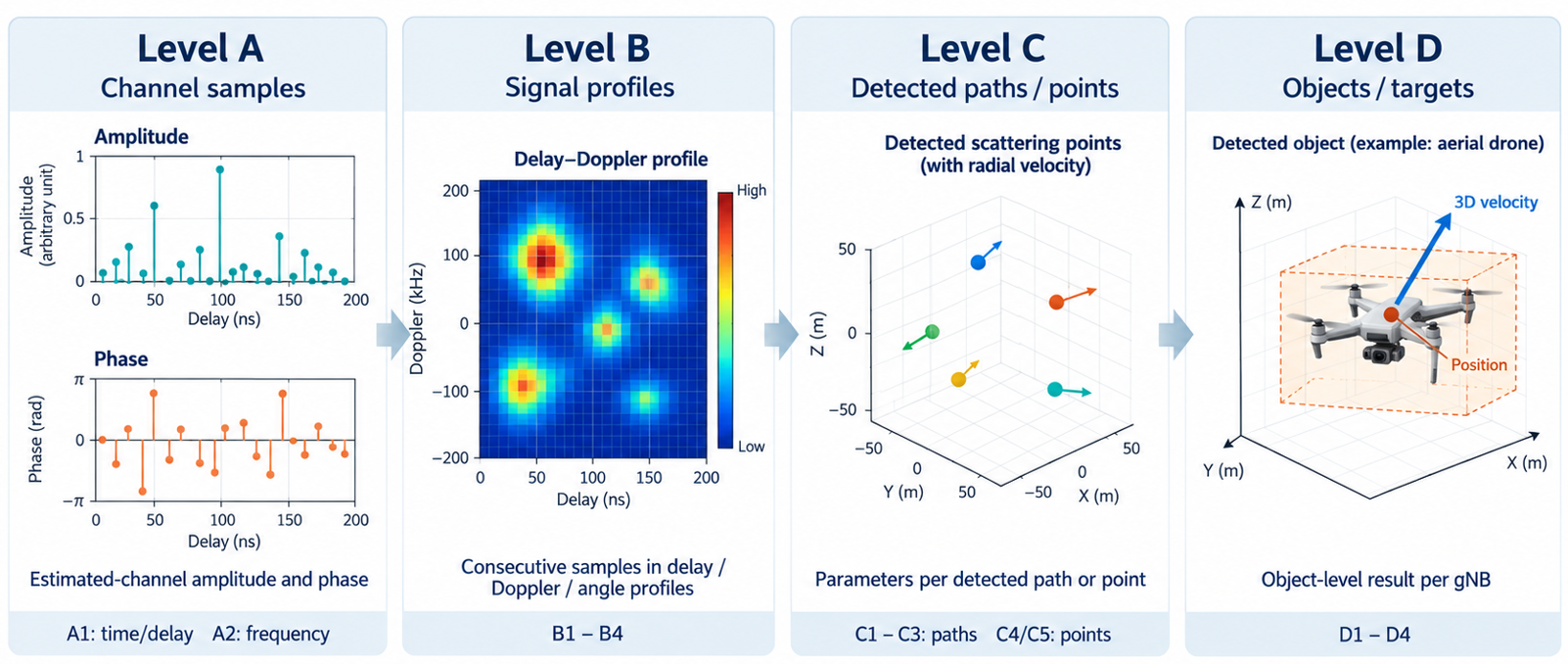}
\caption{Conceptual illustration of the four sensing-information reporting levels considered for gNB--SenF information transfer. Level~A conveys channel samples, Level~B conveys signal profiles, Level~C conveys detected path- or point-level parameters, and Level~D conveys object-level information. The arrows indicate increasing information abstraction rather than a mandatory sequential reporting chain.}
\label{fig:comst_data_levels}
\end{figure*}

The general processing hierarchy described in Section~III-B is instantiated in
the Release~20 study through four gNB--SenF reporting levels, summarized in
Fig.~\ref{fig:comst_data_levels}. Level~A transfers channel-like amplitude and phase information, while Level~D transfers object-level states. Levels~B and~C provide intermediate profile- and parameter-level representations. Moving from Level~A to Level~D reduces transport and centralized processing requirements while increasing the amount of local processing performed at the gNB.

Level~C emerged as a practical baseline because it preserves information useful for multi-gNB fusion without transporting raw samples or profiles~\cite{tr38765,meng_2025_wc}. Level~D is attractive when transport is constrained or the gNB already performs reliable object formation~\cite{tr38765}. However, object-level reports embed local detector thresholds, association rules, coordinate transformations, and uncertainty models. Interoperable fusion therefore requires common semantics, reference coordinates, timing, confidence definitions, and quantization rather than only a common message container~\cite{hall1997multisensor,meng_2025_wc}.

\subsubsection{Service Architecture and Procedures}
\label{subsubsec:service_architecture}

From the 5G system architecture perspective, TR~23.700-14 establishes the Release~20 Stage~2 study baseline, which is being developed normatively in TS~23.137~\cite{tr2370014,ts23137}. The scope is deliberately constrained: the gNB is the sensing entity, aerial-object detection and tracking is the initial service, and a newly introduced SenF coordinates the task. The SenF can be functionally separated into sensing control (SenF-C) and sensing processing (SenF-P), although their internal interface is not standardized in the initial baseline.

The architecture in Fig.~\ref{fig:comst_architecture} connects the SenF to the RAN and to sensing-service consumers. An Application Function (AF) without a trusted relationship with the operator accesses the sensing service through the Network Exposure Function (NEF). N33 connects the AF and NEF, while SEN1 connects the NEF and SenF. An operator-trusted AF can interact directly with the SenF through SEN3.

A sensing service begins when an AF establishes a subscription for sensing-result notifications. After the request is authorized and executed, the results are delivered according to the configured one-time, periodic, or event-triggered reporting mode. The NEF authorizes an untrusted AF and discovers a suitable SenF. The SenF authorizes the request, interprets the external target sensing area and performance parameters, selects suitable sensing entities, and obtains sensing data or results. The SenF then generates or consolidates the result and returns it through the notification path. Modification and unsubscription procedures update or terminate the task.

The request can include a sensing time window, target area, reporting parameters, positioning and velocity accuracy, missed-detection and false-alarm requirements, and service-specific conditions. The external target area uses the geographical-area description framework of TS~23.032, including horizontal shape and altitude information~\cite{ts23032}. This mapping connects Stage~1 service intent to SenF authorization, sensing-entity selection, RAN configuration, result generation, and exposure.

The Release~20 baseline has a deliberately limited scope. It assumes gNB-based sensing within a single Public Land Mobile Network (PLMN), assigns one SenF to each request, and omits a dedicated sensing-data storage function. TS~23.137 should therefore be characterized as a draft Release~20 Stage~2 architecture specification for cellular sensing services rather than a comprehensive end-to-end architecture covering all sensing modes and sensing-entity types.

\subsection{From the Release~20 Baseline Toward 6G}

\subsubsection{Prospects for NR ISAC Normative Work}

The approval of the Release~20 NR ISAC work item marks the first focused 3GPP standardization step toward interoperable NR sensing support in 5G-Advanced~\cite{isac_5ga_wid}. Its scope remains centered on gNB-based monostatic sensing for UAV targets, with no general UE impact. This limited baseline reduces implementation and coordination risk while allowing the Ns interface, NsAP procedures, reporting formats, and security treatment to mature.

The first normative priority is a stable gNB--SenF relationship. NsAP must define interface setup, configuration update, sensing initiation, modification, abort, failure handling, and measurement reporting together with transaction identity, cause values, timer behavior, transport, and error recovery. The gNB split architecture also requires a clear internal allocation of responsibility. The Ns endpoint is associated with the gNB-CU, while the gNB-DU performs time-critical radio operation over the F1 relationship. Intra-gNB TRP--TRP sensing may extend the monostatic baseline when it can be controlled without introducing inter-gNB architecture impact~\cite{3gpp_38_401,isac_5ga_wid}.

The second priority is consistent measurement representation and interpretation across gNBs and the SenF. The work item focuses on Levels~C and~D and on quantization for the UAV use case. Path-level delay, Doppler, angle, received power, and confidence require reference units, coordinate systems, dynamic ranges, invalid-value handling, and timing. Object-level reports of position, velocity, identity, and confidence additionally require common rules for target association and for track initiation, maintenance, and termination. A compact report is useful only when different vendors interpret the same bits as the same physical quantity and uncertainty.

Reporting control is also expected to mature beyond the basic request--response sequence. One-time, periodic, and event-triggered reporting require unambiguous conditions, hysteresis, confidence and freshness rules, and behavior when the requested quality cannot be sustained. Controlled interface removal and state cleanup would prevent stale sensing configurations. Security and privacy require cooperation with SA3 because even non-cooperative targets can reveal sensitive location, behavior, or environmental information. Authorization, integrity, confidentiality, auditability, and exposure policy must cover both the measurements and the derived object results.

These prospects do not imply that every Release~20 study option will become normative. A study compares architectural and protocol alternatives, whereas the normative phase selects a tractable subset. Accordingly, we treat the work item as a bridge from the current baseline to interoperable implementation rather than as a complete ISAC specification.

\subsubsection{Vision and Future Directions for 6G ISAC}

The 6G direction broadens the target classes, sensing modes, frequency ranges, and network roles beyond the constrained NR baseline. The following items represent study directions and architectural questions rather than committed normative features. They combine the RAN-facing implications of expanded sensing with the ongoing Stage~2 architecture study in TR~23.801-01~\cite{tr2380101}.

\begin{itemize}

\item \textbf{Expanded sensing modes and target classes:}
Future evaluation must move beyond gNB monostatic UAV sensing toward TRP--TRP, TRP--UE, UE--TRP, UE--UE, and UE-monostatic operation, including combinations of these modes. Humans, vehicles, AGVs, infrastructure, and small hazards have different RCS distributions, mobility characteristics, micro-Doppler signatures, and clutter overlap. Ground targets may not be separable from the environment through radial velocity alone. Range-profile evolution, spatial discrimination, micro-motion, and multi-node geometry therefore become more important. A CPI must also remain valid under oscillator error, scheduling gaps, node motion, and target dynamics.

\item \textbf{Waveforms and sensing reference signals:}
Communication reference signals (RSs) offer reuse and backward compatibility, but their comb pattern, occupied bandwidth, symbol placement, and coherent duration may not satisfy every range and velocity requirement. Candidate directions include denser or modified communication-RS patterns, time-staggered allocations, non-uniform resource mapping, and a dedicated sensing RS when enhanced communication signals are insufficient. The waveform and RS cannot be evaluated separately because the resource pattern determines the effective sensing bandwidth and coherent observation time. Peak-to-average power ratio (PAPR), cyclic prefix, subcarrier spacing, and phase continuity also affect self-interference cancellation, ambiguity, and weak-target detection~\cite{wei_2023_iotj,gonzalez_2024_pieee}.

\item \textbf{Multi-node architecture and UE participation:}
The 6G architecture study extends the SenF concept toward RAN- and UE-based sensing entities and multiple monostatic and bistatic modes~\cite{tr2380101}. Multi-node sensing requires entity discovery, transmitter--receiver pairing, synchronization, measurement association, fusion, result ownership, and service continuity. The architecture must determine whether an anchor access node, centralized SenF, distributed processing function, or processing hierarchy combines the data. UE participation further introduces capability registration, consent, authorization, mobility notification, compensation, context transfer, and sensing-entity reselection. Multi-SenF coordination, storage and reuse, common time references, and non-3GPP sensor ingestion remain open architectural dimensions.

\item \textbf{Radio-resource coordination and interference handling:}
Future standardization may need to provide configurable sensing resources across time, frequency, space, and power, together with the measurements and signaling required for interference coordination. These mechanisms should accommodate the different resource requirements of initial detection and subsequent tracking. They should also address monostatic self-interference, inter-cell sensing interference, and interference between sensing reference signals and communication channels. The detailed scheduling, interference-cancellation, and target-aware beam-selection algorithms can remain implementation specific.

\end{itemize}

The resulting 6G problem is inherently cross-layer. Waveforms require calibrated target and clutter models, fusion requires timing, association, and consistent report interpretation, and sensing estimates require timely authorization, transport, and processing. These dependencies motivate the simulation, measurement, and validation considerations examined in the following sections.

%% file: figures/ns_interface_corrected.tex
\begin{tikzpicture}
  \node[anchor=south west,inner sep=0] (art) at (0,0)
    {\includegraphics[width=\columnwidth]{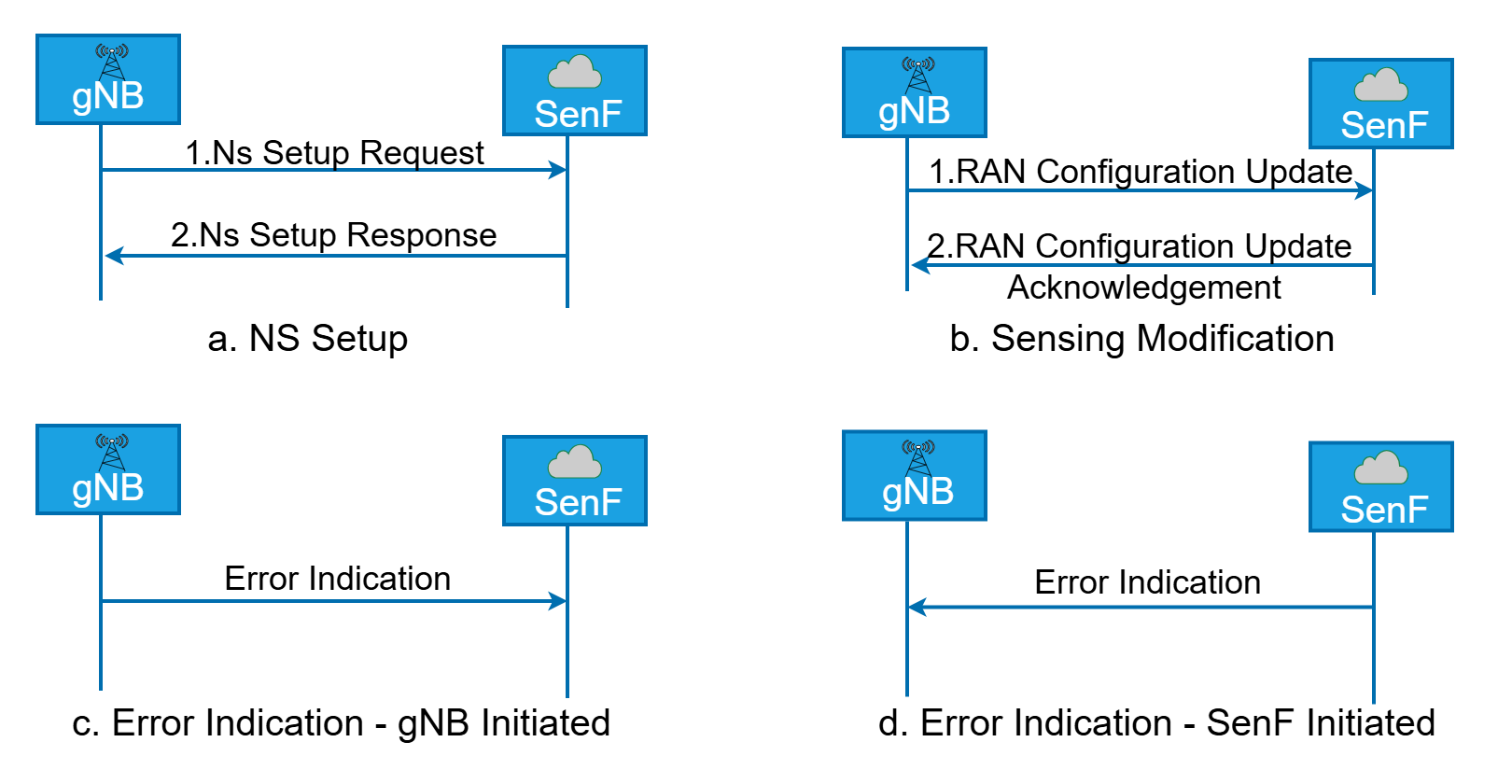}};
  \begin{scope}[x={(art.south east)},y={(art.north west)}]
    \fill[white] (0.52,0.50) rectangle (1,0.59);
    \node[font=\sffamily\fontsize{6.2}{7.5}\selectfont,inner sep=0]
      at (0.765,0.565) {b. RAN Configuration Update};
  \end{scope}
\end{tikzpicture}

%% file: sections/05_simulation.tex
\section{3GPP-Aligned ISAC Simulation: Methodology, Case Studies, and Results}
\label{sec:simulation}

This section reviews simulation methodologies for cellular ISAC and presents two case studies with original performance results. The review covers link-level sensing, system-level communication evaluation, multi-TRP sensing, standards alignment, and reproducibility. The first case study evaluates multi-TRP UAV detection and state estimation at the multi-link link level. The second evaluates sensing-assisted predictive handover for a mobile UE at the system level. Both studies use scenarios, channel models, antenna or link assumptions, and radio parameters drawn from relevant 3GPP documents.

\subsection{Review of ISAC Simulation Methods and Evaluation Levels}

Existing ISAC simulation studies span waveform- and link-level modeling, network-scale system-level evaluation, and multi-node sensing frameworks. These evaluation levels differ in the physical detail they preserve, the spatial and temporal scale they represent, and the performance metrics they can meaningfully support. We review the literature according to its channel and signal abstractions, network scope, sensing-processing functions, and reported performance indicators.

\subsubsection{Link-Level Sensing Evaluation}

Link-level sensing simulation represents the signal chain that produces observations for target detection and parameter estimation. A typical implementation begins by generating the waveform and resource grid and applying the transmit and receive array responses. The signal is then propagated through the target and background channels with receiver impairments and interference. Receiver processing forms delay--Doppler--angle representations for detection, association, and state estimation. This abstraction supports controlled evaluation of waveform ambiguity, coherent processing gain, and angular resolution. It also captures target RCS, polarization, clutter cancellation, self-interference, and estimation accuracy~\cite{liu_2023_spm,liu_2022_jsac,wei2025isacchannel}.

The required detail is higher than that of a communication-only link. A
communication receiver may represent multipath mainly through its effect on
channel estimation and decoding. A sensing receiver must preserve the delays,
angles, Doppler shifts, phases, and time evolution that encode target state.
For bistatic or multistatic sensing, it must also preserve the two propagation
legs, node geometry, target-dependent scattering, and a common time and
frequency reference. Evaluating array-dependent performance additionally requires explicit modeling of element locations, radiation patterns, polarization slants, and the target polarization transfer matrix.

Dai \emph{et al.} provide a complementary MIMO-OFDM framework for spatial, delay, and Doppler estimation in far- and near-field conditions~\cite{dai2026mimoofdmisac}.

Link-level studies should apply the detection, false-alarm, association, and
conditional-estimation conventions defined in
Section~\ref{subsec:evaluation_metrics}. Beyond these common definitions,
theoretical bounds can provide useful performance references, but they do not
replace the operating characteristics of an implemented detector or the
finite-sample error distribution of an estimator
~\cite{liu_2022_cst,zhang_2021_jstsp}.

\subsubsection{System-Level ISAC Evaluation}

System-level simulation represents large networks and extended time periods using simplified waveform and receiver models. It includes multiple sites, UEs, targets, resource intervals, and mobility events. Correlated large- and small-scale channel states are generated to model interference, scheduling, association, handover, and service-level performance. Physical-layer behavior is commonly represented through calibrated mappings from the signal-to-interference-plus-noise ratio (SINR) to the block error rate (BLER), link-adaptation curves, detection-probability tables, and measurement-error models.

Stochastic-geometry analysis complements these simulations by representing random node locations and characterizing spatially averaged sensing and communication performance~\cite{armeniakos2026stochastic}.

An ISAC system-level abstraction requires more than inserting a position-error variance into an existing network simulator. Sensing information has an update period, age, uncertainty, missed-detection process, and possibly correlated biases. A network decision also has preparation, execution, expiry, and recovery states. Link-to-system calibration should therefore preserve the failure mechanisms that affect the performance metric under study. For example, a predictive-handover study must distinguish the data-plane MCS from the more robust MCS used for handover execution and recovery. Otherwise, an aggressive data-rate setting can unintentionally determine the handover-failure rate.

The appropriate simulation abstraction depends on the evaluation objective. Resource-allocation performance can often be studied using analytical sensing-SINR constraints, whereas mobility evaluation requires time-evolving measurements and a protocol state machine. For transparent interpretation, the simulator should distinguish parameters derived from standards, quantities calibrated through link-level evaluation, and study-specific policy assumptions.

\subsubsection{Networked and Multi-TRP Sensing}

Networked sensing provides observation diversity while introducing
synchronization, association, transport, and fusion requirements. A simulation
should explicitly implement one of the processing and reporting interfaces
introduced in Section~III-B and represented by the Levels~A--D framework in
Fig.~\ref{fig:comst_data_levels}. It should not transfer perfect target states between nodes unless an ideal reference method is being evaluated.

Sample- or profile-level exchange requires explicit timing, frequency, phase,
and array calibration. Parameter- or object-level exchange instead requires
models for local detection, uncertainty, association, missing reports, and
transport latency. In every case, the simulation should retain correlated
visibility, clutter, synchronization error, and common illumination effects
across sensing links~\cite{li_2024_twc,meng_2025_wc,meng_2025_twc,han2025network}.

The existing literature consequently combines complementary simulation tools rather than relying on a universal ISAC simulator. Geometry-based stochastic channel models provide repeatable network realizations and spatial consistency. Deterministic ray tracing captures site-specific propagation and supports environment-specific calibration. Waveform-level simulators represent signal processing for target detection and parameter estimation, whereas radio system-level simulators model scheduling, interference, and mobility. Hardware-in-the-loop and channel-emulation platforms can further reveal processing latency and implementation impairments~\cite{zhang2023channel,wei2025isacchannel,pedersen2024isacsls}.

\subsection{Standards Alignment and Evaluation Protocol}

The two case studies adopt 3GPP scenarios, channel models, and radio parameters, with study-specific geometry, processing, and control assumptions identified separately. Table~\ref{tab:section_iv_parameters} summarizes their principal configurations and metrics. Methods share matched realizations, and calibration data are separated from the reported evaluation. Detection and false-alarm probabilities accompany conditional estimation CDFs, while handover outcomes accompany the user-throughput CDF.

In the communication case, reference signal received power (RSRP) guides mobility decisions. Link calibration uses a single-input single-output (SISO) configuration, with frequency-domain resources specified in physical resource blocks (PRBs). Radio configurations use the subcarrier spacing (SCS) listed in the table.

\input{tables/evaluation_parameters.tex}

\subsection{Case Study A: Multi-TRP UAV Sensing}
\label{subsec:comst_uav_case}

\subsubsection{Scenario, Channel, and Compared Fusion Methods}

The UAV case uses selected parameters from TR~38.765 Table~6.2-1 Baseline~1, including a 4.9~GHz carrier, 52~dBm maximum BS power, and five UAVs at heights of 25--300~m~\cite{tr38765}. Three TRPs form a study-specific triangular geometry with 500~m pairwise separation. The dual-polarized panels and TR~38.901 target and background channels represent array, RCS, Doppler, and polarization effects~\cite{ref11}. Target-directed observations use 100~MHz system bandwidth and a 32~ms CPI. Blackman--Harris windowing and static-background suppression precede detection~\cite{harris1978windows}. Residual self-interference is included, while inter-TRP interference is omitted. Fusion uses angle estimates extracted from the simulated receiver responses
rather than ground-truth target directions.

Three fusion methods are compared. \emph{Hard-decision fusion} requires consistent hypotheses from at least two TRPs. \emph{Soft-decision fusion} combines threshold-normalized response maps around
candidate positions derived from receiver-estimated angles. \emph{Hybrid fusion} retains accepted hard-decision hypotheses and supplements them with nonduplicate soft-decision hypotheses. All methods apply one-to-one association between reported objects and true targets.

Method-specific thresholds are calibrated against a common 1\% Type-1 FAP budget and frozen before evaluation. The reported descriptive results combine independently generated layouts containing 100 target-present scenes, 500 target opportunities, and 784 target-absent scenes. Type-2 FAP follows the scene-level definition in Section~\ref{sec:standardization}. Position- and velocity-error CDFs include only correctly detected and associated targets. This controlled evaluation illustrates fusion tradeoffs under target-directed acquisition rather than blind spatial search or continuous tracking.

\subsubsection{Multi-TRP Detection and Estimation Results}

\begin{figure*}[!t]
\centering
\includegraphics[width=0.95\textwidth]{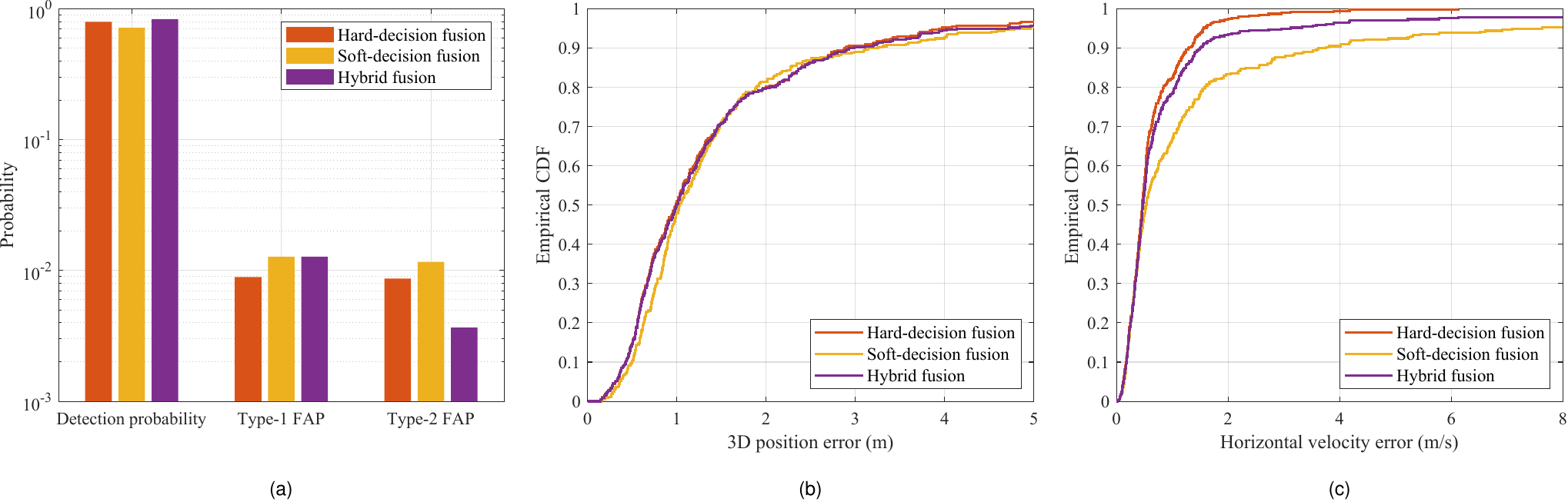}
\caption{Multi-TRP UAV sensing with selected TR~38.765 Baseline-1 parameters and fusion
based on receiver-estimated angles. Frozen policies share a 1\% calibration Type-1 FAP budget. Cumulative evaluation uses 100 target-present scenes (500 targets) and 784 target-absent scenes. (a) Detection probability and Type-1/Type-2 FAP. (b) Conditional 3D position-error CDF. (c) Conditional horizontal-velocity-error CDF. The CDFs include all associated errors without smoothing. Panels (b) and (c) show restricted horizontal ranges without renormalization. Achieved evaluation FAPs need not be identical.}
\label{fig:comst_uav_results}
\end{figure*}

Fig.~\ref{fig:comst_uav_results}(a) reports the cumulative evaluation at a common calibration budget, rather than identical achieved evaluation FAPs. Hard-decision fusion, Soft-decision fusion, and Hybrid fusion detect 398, 359, and 418 of 500 targets, giving detection probabilities of 0.796, 0.718, and 0.836. Their Type-1 FAPs are 7/784 (0.893\%), 10/784 (1.276\%), and 10/784 (1.276\%). Their Type-2 FAPs are 0.867\%, 1.167\%, and 0.367\%, respectively. All 100 target-present scenes are nonempty for each method. These Type-2 values are means of scene-level fractions, not ratios of pooled false reports to pooled reports. Every plotted probability is an observed estimate, with no artificial positive floor.

Hybrid fusion detects four percentage points more targets than Hard-decision fusion in the combined evaluation. A separately seeded extension showed the same ordering. The false-alarm differences remain inconclusive because only a small number of events is observed.

Fig.~\ref{fig:comst_uav_results}(b) and (c) show unsmoothed conditional empirical CDFs with 398, 359, and 418 associated targets. The 90th-percentile 3D position errors are 2.897, 3.260, and 2.986~m for Hard-decision fusion, Soft-decision fusion, and Hybrid fusion. The corresponding horizontal-velocity errors are 1.302, 3.756, and 1.513~m/s. Hybrid fusion therefore improves detection while retaining estimation accuracy close to Hard-decision fusion. These study-specific 3D position and horizontal-velocity metrics should be distinguished from individual horizontal, vertical, or radial metrics in standards evaluations. In particular, the missed-detection probabilities remain above the 5\% reference objective in Table~\ref{tab:comst_tr38765_baseline}, so the results do not establish simultaneous satisfaction of all objectives.

\subsubsection{Key Findings and Evaluation Scope}

Hybrid fusion provides the highest detection probability while retaining conditional position and velocity accuracy close to Hard-decision fusion. Its lower observed Type-2 FAP is descriptive rather than a statistically established false-alarm advantage. These results distinguish detection improvement from estimation accuracy and false-alarm reliability.

These findings apply to a controlled multi-link link-level evaluation with study-specific TRP geometry, target-directed acquisition, measured-angle processing, static-background suppression, and independent layouts. Continuous tracking, blind spatial search, and communication--sensing coexistence are outside the present scope. Conditional estimation CDFs should be interpreted together with detection and false-alarm probabilities.

\subsection{Case Study B: Sensing-Assisted Predictive Handover}
\label{subsec:comst_handover_case}

Li et al. study sensing-assisted handover through position- and maneuver-based triggering, beamforming, and NR frame and protocol design~\cite{li_look_before_switch}. This case study instead evaluates predictive preparation under a common execution policy, with explicit sensing uncertainty, information age, and separate control- and data-plane reliability models.

\subsubsection{Scenario, Mobility, and Compared Handover Methods}

The mobility case considers a UE moving at 60~km/h across a cell boundary in a three-TRP outdoor UMa deployment with 500~m inter-site distance. TR~38.901 propagation and study-specific spatially consistent channel evolution are represented through sector-equivalent scalar links. \emph{Conventional} handover uses current filtered RSRP without predictive preparation. \emph{Ideal} uses future filtered RSRP as a noncausal upper reference. \emph{Practical} uses constant-velocity prediction from position and velocity estimates with standard deviations of 15~m and 1.5~m/s, respectively, and sensing-information age uniformly distributed over 0.1--0.5~s.

All methods share the same execution condition, requiring a 3~dB RSRP advantage for a study-specific 3~s persistence interval. This interval is distinct from the standardized TimeToTrigger values in TS~38.331~\cite{3gpp_ts_38331}. Predictive preparation uses a 2~s look-ahead window and reduces the modeled access time from 60 to 30~ms when successful. The interruption benefit therefore depends on this assumed reduction in access overhead. The primary model uses robust MCS~4 for control and adaptive data-plane MCS selection targeting 10\% BLER. Link calibration follows coded NR Physical Downlink Shared Channel (PDSCH) and Downlink Shared Channel (DL-SCH) processing and the relevant NR resource and transport-block specifications~\cite{3gpp_ts_38104,3gpp_ts_38211,3gpp_ts_38212,3gpp_ts_38214}. A coupled MCS~9 sensitivity case illustrates the effect of using the same MCS for data and handover execution.

The evaluation uses 512 paired drops spanning representative LOS and NLOS conditions. Handover failure rate is the fraction of attempts that fail target-access qualification within the 120~ms execution timer. Interruption extends from attempt initiation to the first successfully decoded data slot, including any recovery, and averages all attempts with observed data resumption. User throughput is the successfully decoded bit rate averaged over the 0.5~s interval following each attempt, with unavailable slots and failed blocks contributing zero. Its CDF is conditioned on successful handover.

\subsubsection{Handover and User-Throughput Results}

\begin{figure}[!t]
\centering
\includegraphics[width=0.95\columnwidth]{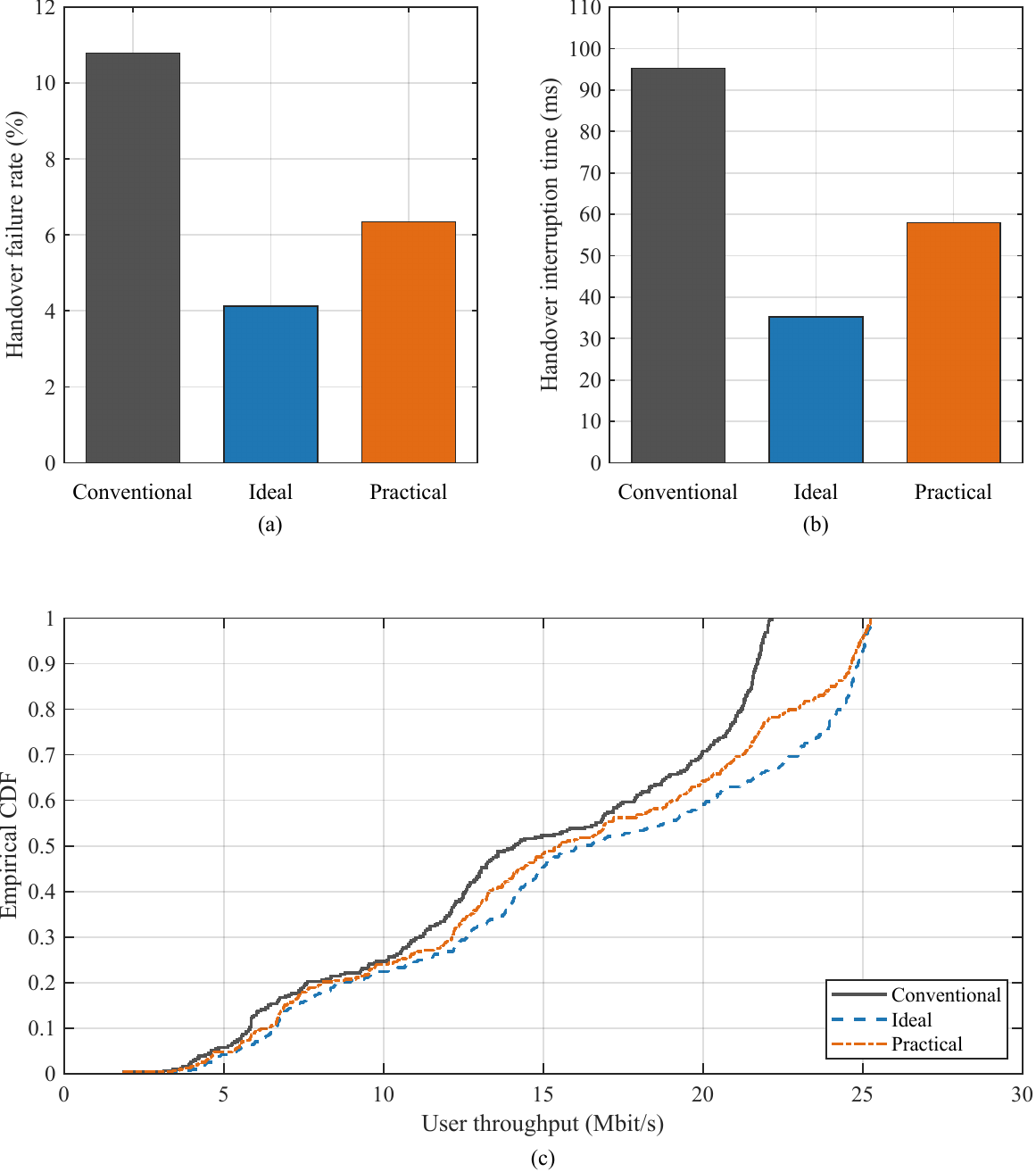}
\caption{Predictive-handover results from the 512-drop paired evaluation.
(a) Handover-failure rate under coupled MCS~9 operation.
(b) Mean handover interruption time under adaptive MCS selection with
separate control- and data-plane reliability.
(c) Empirical CDF of user throughput under the same primary configuration.
Ideal predictive sensing-assisted handover provides a noncausal upper reference, while Practical
predictive sensing-assisted handover uses noisy and delayed mobility information.}
\label{fig:comst_handover_results}
\end{figure}

Fig.~\ref{fig:comst_handover_results} summarizes the results. Under the primary configuration with separate control- and data-plane reliability, each method encounters 313 handover attempts. Conventional experiences one failure (0.32\%), whereas Ideal and Practical experience none. The small number of failures prevents a general conclusion about handover reliability.

The mean handover interruption times are 95.3, 35.2, and 58.0~ms for
Conventional, Ideal, and Practical, respectively. Their mean user throughputs
are 14.50, 16.59, and 15.79~Mbit/s. Relative to Conventional, Practical reduces
interruption by 37.3~ms and increases throughput by 1.29~Mbit/s. Ideal
provides a further improvement over Practical, showing the effect of sensing
uncertainty and information age.

The empirical throughput CDF is based on successful handovers and is not smoothed or interpolated.

Fig.~\ref{fig:comst_handover_results}(a) presents a sensitivity evaluation in
which MCS~9 is applied jointly to data transmission and handover execution.
The handover-failure rates are 0.108, 0.041, and 0.063 for Conventional, Ideal,
and Practical, respectively. The increased failure rate demonstrates why an
aggressive data MCS should not automatically determine handover-access
reliability.

\subsubsection{Key Findings and Evaluation Scope}

The results demonstrate reduced handover interruption and improved user throughput near the handover event, while the small number of failures prevents a general conclusion about handover reliability. Separating control- and data-plane link adaptation clarifies the mechanism. Predictive sensing improves target preparation, while robust control-plane transmission maintains reliable execution. The difference between Ideal and Practical reflects the effects of sensing uncertainty and information age.

The evaluation uses study-specific spatially consistent channel evolution, UE mobility, and sector-equivalent scalar links. Link calibration assumes ideal synchronization and channel knowledge, while handover procedures are represented through timers and BLER-driven state transitions. The throughput CDF is conditioned on successful handover, and Ideal provides a noncausal upper reference. A broader reliability comparison would require a larger independent evaluation.

\subsection{Cross-Case Findings}

The two case studies operate at different evaluation levels but share one
methodological requirement. Sensing information must be evaluated at the
interface where it changes a decision. In the UAV case, this interface is the fusion and association rule. A common calibration false-alarm budget reveals differences in detection, unassociated reports, and velocity-error tails. In the mobility case, the relevant interface is handover preparation. Its benefit depends on sensing uncertainty, information age, preparation timing, expiry, and control-link reliability.

Together, the cases show that an ISAC gain depends on both the sensing output and the decision process that use it. The two cases illustrate different interfaces between sensing and network operation. In the UAV case, threshold calibration and fusion determine the tradeoff among detection, false alarms, and conditional estimation accuracy.
In the mobility case, sensing uncertainty and information age affect
predictive preparation, while the separation of control- and data-plane
reliability determines whether data-link adaptation obscures the handover
mechanism.

\FloatBarrier

%% file: tables/evaluation_parameters.tex
\begin{table*}[!t]
\centering
\caption{Principal Configurations of the Two Simulation Case Studies}
\label{tab:section_iv_parameters}
\footnotesize
\setlength{\tabcolsep}{4pt}
\begin{tabular}{p{0.18\textwidth}p{0.37\textwidth}p{0.37\textwidth}}
\toprule
\textbf{Item} & \textbf{UAV sensing} & \textbf{Predictive handover} \\
\midrule
Evaluation level and basis & Multi-link link level. TR~38.765 Baseline~1 and TR~38.901 sensing channel & System level. TR~38.901 UMa and NR link calibration using TS~38.104/38.211/38.212/38.214 \\
Deployment & Three TRPs, 500~m pairwise separation. Five UAVs at 25--300~m height & Three TRPs, 500~m inter-site distance. One UE moving at 60~km/h \\
Radio configuration & 4.9~GHz, 52~dBm maximum BS power, 100~MHz bandwidth, 30~kHz SCS, 32~ms CPI & 6~GHz, 20~MHz bandwidth, 30~kHz SCS, 51 PRBs \\
Channel and antenna & Target/background channels with RCS and polarization. Dual-polarized $(8,8,2,1,1;4,8)$ panel & Spatially consistent UMa links with sector-equivalent scalar antennas and coded SISO link calibration \\
Compared methods & Hard-decision, soft-decision, and hybrid fusion & Conventional handover, Ideal and Practical predictive handover \\
Evaluation size & 100 target-present scenes (500 targets) and 784 target-absent scenes & 512 paired drops spanning LOS/NLOS conditions \\
Performance metrics & Detection, Type-1/Type-2 FAP, conditional position- and velocity-error CDFs & Handover failure rate, interruption time, and user-throughput CDF \\
\bottomrule
\end{tabular}
\end{table*}

%% file: sections/06_testbeds.tex
\section{ISAC Prototypes, Testbeds, Measurements, and Experimental Case Study}
\label{sec:testbeds}

Simulation can evaluate an ISAC mechanism under controlled conditions, but it
cannot by itself establish whether the assumed waveform, processing latency,
radio-frequency dynamic range, or target response is physically realizable.
Prototypes and testbeds therefore validate the implementation chain, whereas
channel and target measurements determine whether the propagation and
scattering assumptions used in simulation are physically representative. We
first review these complementary experimental roles and then present a UAV RCS
measurement case study. The measurements connect target visibility to the
TR~38.901 Clause~7.9 channel abstraction discussed in
Sections~\ref{sec:foundations} and~\ref{sec:standardization}.

\subsection{Prototype and Testbed Literature}

\subsubsection{Waveform and Signal-Processing Prototypes}

ISAC prototypes commonly begin with a shared radio front end and an OFDM or multicarrier waveform. The
communication branch performs synchronization, channel estimation,
equalization, and decoding, whereas the sensing branch forms
delay--Doppler--angle statistics and then performs detection or estimation.
Software-defined radios provide implementation flexibility, while
field-programmable gate arrays (FPGAs), graphics processing units (GPUs), or
dedicated accelerators are used when coherent processing and beam scanning must
run in real time. Recent surveys show a progression from single-link
demonstrations toward MIMO, near-field, and
networked experiments~\cite{zhang_2022_cst,luo2026isacsurvey,zhang2026isacevolution}.

A reliable prototype should report both communication and sensing functions
under a common operating configuration~\cite{zhang_2022_cst,liu_2022_jsac,luo2026isacsurvey}.
Communication performance may be represented by error vector magnitude (EVM),
BLER, throughput, latency, or coverage. Sensing performance may be represented
by detection probability, FAP, range, velocity, and angle error, resolution,
or tracking continuity. Reporting the best communication and sensing results
from different resource allocations does not demonstrate simultaneous
operation. The occupied bandwidth, coherent processing interval, array
aperture, transmit power, self-interference conditions, processing latency,
and ground-truth acquisition procedure should therefore be documented with the
reported results~\cite{wang2024experimental}.

Prototype-based validation usually covers a narrower range of operating
conditions than can be examined through simulation~\cite{luo2026isacsurvey}.
A laboratory demonstration may verify the complete baseband processing chain,
but only for a particular environment, geometry, target, and interference
condition. Repeatable calibration and access to intermediate data are therefore
valuable~\cite{zhang2024isacchannel,wei2025isacchannel}. Raw or calibrated
channel responses allow subsequent algorithms to be compared without repeating
the over-the-air measurements. Recorded timing traces also indicate whether the
sensing information is delivered promptly enough to support network
control~\cite{11028899}.

\subsubsection{Networked ISAC Demonstrations}

Networked demonstrations experimentally evaluate the processing and reporting
splits introduced in Section~III-B and Fig.~\ref{fig:comst_data_levels}. Their
main purpose is to determine whether the selected representation preserves
sufficient sensing information under practical synchronization, calibration,
transport-capacity, processing-latency, and packet-delivery constraints.
Higher-level reports should also retain the effects of local thresholding,
association, uncertainty estimation, and track management.

Open and disaggregated RAN testbeds are useful for evaluating these processing
splits because radio measurements, near-real-time control, and application
logic can be placed at different nodes~\cite{11028899}. Recent prototypes have
demonstrated monostatic, bistatic, and network sensing with coordinated radio
nodes, multimodal synchronization, and sensing-assisted
communication~\cite{yang2025isacprototype,i2025cooperativeisac}. For cellular
ISAC, however, multi-node radio operation alone does not validate a complete
end-to-end sensing service. Such validation must additionally cover
sensing-task authorization and management, standardized information transfer,
result delivery, processing and transport latency, and failure handling.

Two bistatic prototypes illustrate the synchronization and calibration requirements. Brunner \emph{et al.} validate over-the-air synchronization for OFDM-based ISAC and examine how residual synchronization errors and decoding failures affect sensing and communication~\cite{brunner2025bistatic}. Giroto \emph{et al.} demonstrate a $4\times8$ MIMO-OFDM setup at 27.5~GHz with distributed synchronization, transmitter predistortion, and receiver calibration~\cite{giroto2026bistatic}. These studies provide concrete implementation references for the timing and array assumptions used in multi-node simulation.

These implementation-oriented testbeds establish whether sensing information
can be generated, transported, and consumed under realistic system
constraints. The complementary question is whether the propagation and
target-response assumptions underlying that information are physically
representative; this motivates the measurement literature considered next.

\subsection{Channel and Target Measurement Literature}

ISAC measurement campaigns address two related but distinct objects. A
\emph{propagation-channel} campaign characterizes background delay, angle,
Doppler, path loss, clutter, and the sharing of clusters among communication
and sensing links. A \emph{target-response} campaign characterizes the field
scattered by a specified object as a function of frequency, incidence,
observation direction, polarization, and target state. Combining the two
without preserving this distinction makes it difficult to transfer a result to
a different deployment~\cite{10334037,10494709,wei2025isacchannel}.

Liu \emph{et al.} combine UAV and ground-level measurements at 4~GHz with ray tracing to construct an object-oriented channel model for low-altitude urban spaces~\cite{liu2026object}. Their separation of static background and dynamic-object contributions illustrates how measurements can calibrate the propagation environment used in sensing simulation.

The target-response dependencies introduced in Section~\ref{sec:foundations}
set the requirements for controlled RCS measurement. A campaign needs a known
reference target, background characterization, geometric and polarization
calibration, sufficient frequency sampling for range gating, and a statement
of whether far-field conditions are met~\cite{32062}. Statistical descriptions
are useful when detailed orientation is unknown, whereas deterministic angular
patterns are useful when target attitude is available. Both may be required by
a target-aware channel model~\cite{4102863,unified_rcs_3gpp_isac}.

Prior works on UAV target-response characterization include electromagnetic
simulation, measured monostatic distributions, detection analysis based on
measured RCS, and standardized modeling
contributions~\cite{10064841,9982651,10745726,3rd20243gpp}. These studies
demonstrate that the UAV target response varies significantly with platform
geometry, aspect, and observation angle, but distributed cellular sensing
additionally needs bistatic and polarimetric characterization. Rotating
propellers introduce micro-Doppler and a time-varying response that a static
airframe measurement does not capture~\cite{10548468}. Thus, no single
campaign can supply all channel-model parameters. A useful data set should
specify the target state and the dimensions over which it can be generalized.

The experimental case study that follows deliberately focuses on the second of
these roles: controlled characterization of the target response. It therefore
complements, rather than replaces, propagation-channel measurements.

\subsection{Experimental Case Study: UAV RCS Characterization}
\label{subsec:comst_rcs_case}

Accordingly, this experimental case study uses an RCS measurement testbed to
characterize the dependence of UAV target response on frequency, observation
aspect, platform geometry, and bistatic angle. The study evaluates physical
target scattering rather than end-to-end sensing performance. The resulting
measurements provide empirical inputs for parameterizing and validating UAV
RCS models within the TR~38.901 target-aware channel framework and for
simulations based on the TR~38.765 UAV sensing scenario.

\subsubsection{Measurement Scenario}

The principal campaign characterizes a DJI Phantom~2 UAV. The platform is
approximately 35~cm diagonally and 20~cm high. It is mounted on a turntable and
rotated through a full azimuth scan in $1^\circ$ increments. A Keysight PNA-X
records 2001 frequency points from 11 to 26~GHz using vertically polarized
directional antennas. The measurements are conducted in a controlled indoor
non-anechoic near-field environment. Background subtraction, calibration, and
range gating are used to extract the target-dominated response. In the bistatic
configuration, the transmitter-to-target and target-to-receiver distances are
both 2.6~m, and the bistatic angle is varied from $0^\circ$ to $90^\circ$ in
$15^\circ$ increments.

\begin{figure*}[!t]
\centering
\includegraphics[width=0.80\textwidth]{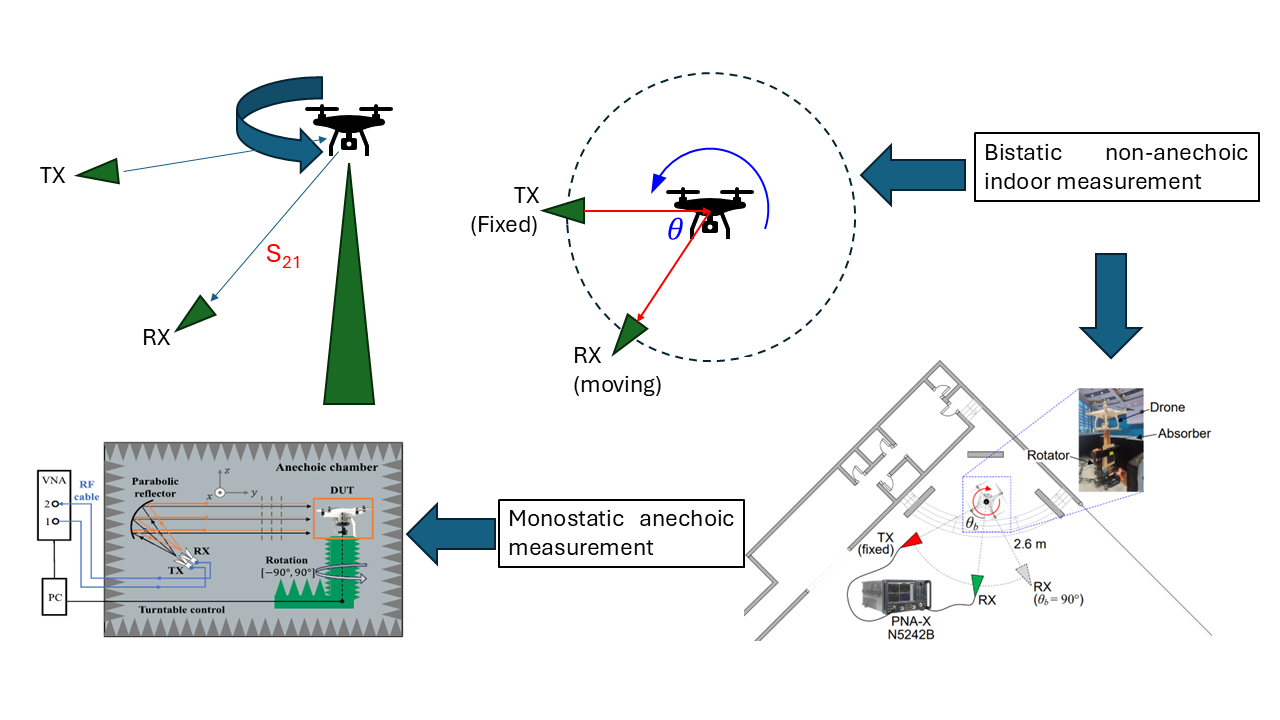}
\caption{UAV RCS measurement configurations and testbeds. The upper
schematics show monostatic and bistatic geometries, including azimuthal target
rotation and receiver displacement. The lower panels show the monostatic
anechoic-chamber and bistatic controlled-indoor setups.}
\label{fig:comst_rcs_setup}
\end{figure*}

The Phantom~2 measurements are complemented by DJI Mini~2, DJI M350~RTK, and
DJI Phantom~3 data sets. Their measurement settings are summarized in
Table~\ref{tab:uav_measurement_settings}. Because the platforms, frequency
ranges, and facilities are not identical, the cross-platform comparison is
descriptive rather than a controlled ranking. The inclusion of bistatic
measurements is particularly relevant to multi-TRP sensing, where monostatic
visibility need not translate to comparable visibility at a spatially
separated receiver. Fig.~\ref{fig:comst_rcs_setup} shows the principal
measurement geometries and testbeds.

\begin{table*}[!t]
\caption{Representative UAV RCS Measurement Settings}
\label{tab:uav_measurement_settings}
\centering
\scriptsize
\setlength{\tabcolsep}{4pt}
\renewcommand{\arraystretch}{1.08}
\begin{tabular}{P{0.14\textwidth} P{0.18\textwidth} P{0.18\textwidth}
P{0.18\textwidth} P{0.18\textwidth}}
\toprule
\textbf{Parameter} &
\textbf{DJI Mini~2} &
\textbf{DJI M350~RTK} &
\textbf{DJI Phantom~2} &
\textbf{DJI Phantom~3} \\
\midrule
Dimensions &
15.9$\times$20.3~cm &
81$\times$67~cm &
35$\times$20~cm &
35$\times$20~cm \\
Frequency band &
21--27~GHz &
21--27~GHz &
11--26~GHz &
1.8--18.2~GHz \\
Frequency points &
501 &
501 &
2001 &
2801 \\
Azimuth sampling &
$0^\circ$--$360^\circ$, $0.5^\circ$ step &
$0^\circ$--$360^\circ$, $0.5^\circ$ step &
$0^\circ$--$360^\circ$, $1^\circ$ step &
$-90^\circ$--$90^\circ$, $2^\circ$ step \\
Measurement regime &
Far field &
Far field &
Near field, 2.6~m &
Far field \\
Environment &
Anechoic &
Anechoic &
Controlled non-anechoic indoor &
Anechoic \\
\bottomrule
\end{tabular}
\end{table*}

\subsubsection{Testbed Implementation and Measurement Processing}

The measurement procedure isolates the UAV-scattered response and converts it
into calibrated RCS samples. Three responses are acquired under the same
geometry. First, an empty-turntable measurement records static environmental
reflections, direct transmitter--receiver coupling, and support-structure
scattering. Second, a metallic calibration cylinder with known 2.8~dBsm RCS is
placed at the target position. Third, the UAV is measured over the required
azimuth and bistatic angles.

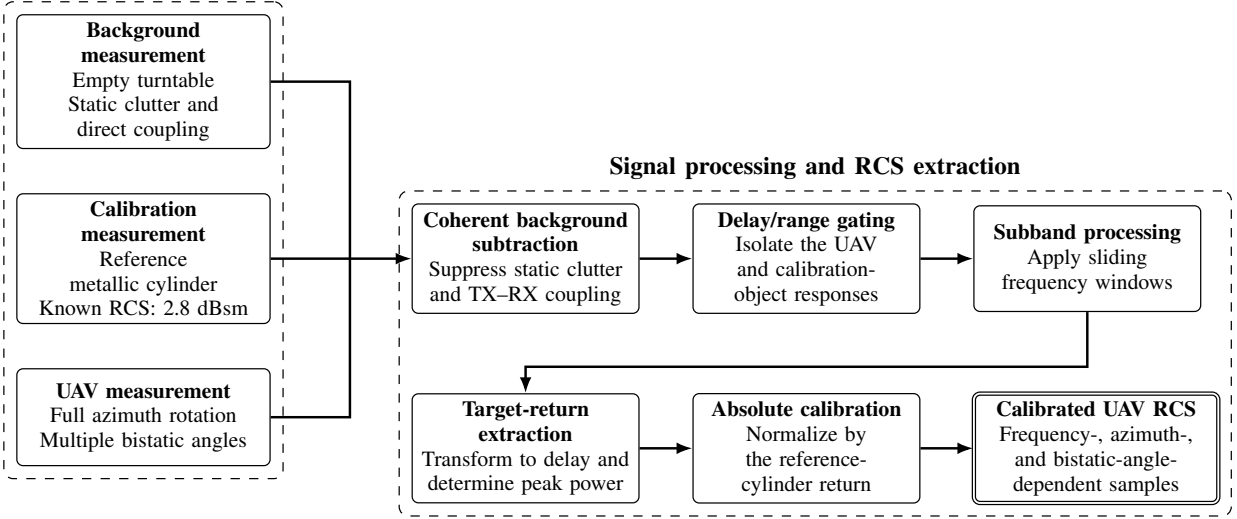
\begin{figure*}[!t]
\centering
\resizebox{0.9\textwidth}{!}{%
\begin{tikzpicture}[
    font=\scriptsize,
    block/.style={rectangle,rounded corners=2pt,draw,align=center,
        minimum height=12mm,text width=24mm,inner sep=2.5pt},
    input/.style={rectangle,rounded corners=2pt,draw,align=center,
        minimum height=11mm,text width=27mm,inner sep=2.5pt},
    output/.style={rectangle,rounded corners=2pt,draw,double,align=center,
        minimum height=12mm,text width=26mm,inner sep=2.5pt},
    arrow/.style={-{Latex[length=1.8mm]},thick},
    group/.style={draw,dashed,rounded corners=3pt,inner sep=4pt}]

\node[input] (bg) {
    \textbf{Background measurement}\\
    Empty turntable\\
    Static clutter and\\
    direct coupling};

\node[input, below=5mm of bg] (cal) {
    \textbf{Calibration measurement}\\
    Reference metallic cylinder\\
    Known RCS: 2.8~dBsm};

\node[input, below=5mm of cal] (dut) {
    \textbf{UAV measurement}\\
    Full azimuth rotation\\
    Multiple bistatic angles};

\node[block, right=16mm of cal] (subtraction) {
    \textbf{Coherent background subtraction}\\
    Suppress static clutter and TX--RX coupling};

\node[block, right=6mm of subtraction] (gating) {
    \textbf{Delay/range gating}\\
    Isolate the UAV and calibration-object responses};

\node[block, right=6mm of gating] (subband) {
    \textbf{Subband processing}\\
    Apply sliding frequency windows};

\node[block, below=9mm of subtraction] (peak) {
    \textbf{Target-return extraction}\\
    Transform to delay and determine peak power};

\node[block, right=6mm of peak] (calibration) {
    \textbf{Absolute calibration}\\
    Normalize by the reference-cylinder return};

\node[output, right=6mm of calibration] (rcs) {
    \textbf{Calibrated UAV RCS}\\
    Frequency-, azimuth-, and bistatic-angle-dependent samples};

\coordinate (merge) at ($(subtraction.west)+(-7mm,0)$);

\draw[thick] (bg.east) -| (merge);
\draw[thick] (cal.east) -- (merge);
\draw[thick] (dut.east) -| (merge);
\draw[arrow] (merge) -- (subtraction.west);
\draw[arrow] (subtraction) -- (gating);
\draw[arrow] (gating) -- (subband);
\draw[arrow] (subband.south) |- ($(peak.north)+(0,3mm)$) -- (peak.north);
\draw[arrow] (peak) -- (calibration);
\draw[arrow] (calibration) -- (rcs);

\node[group,fit=(bg)(cal)(dut),
    label={[font=\footnotesize\bfseries]above:Measurement acquisition}] {};

\node[group,fit=(subtraction)(gating)(subband)(peak)(calibration)(rcs),
    label={[font=\footnotesize\bfseries]above:Signal processing and RCS extraction}] {};

\end{tikzpicture}%
}
\caption{Measurement and processing pipeline for UAV RCS extraction. The
background, reference-cylinder, and UAV responses are recorded under identical
geometries. Coherent subtraction and range gating isolate the target response,
which is processed in subbands and normalized using the known cylinder RCS.}
\label{fig:measurement_pipeline}
\end{figure*}

Fig.~\ref{fig:measurement_pipeline} summarizes the complete acquisition and
processing chain. Complex-valued background subtraction removes static
components coherently rather than only in received power. The result is
transformed to the delay domain, and a range gate is placed around the known
target distance to suppress residual multipath. Processing in smaller
frequency subbands preserves frequency dependence while keeping the target
response compact within each delay-domain gate. Finally, the UAV return is
normalized by the reference-cylinder return to obtain an absolute RCS
estimate.

\begin{figure}[!t]
\centering
\includegraphics[width=\columnwidth]{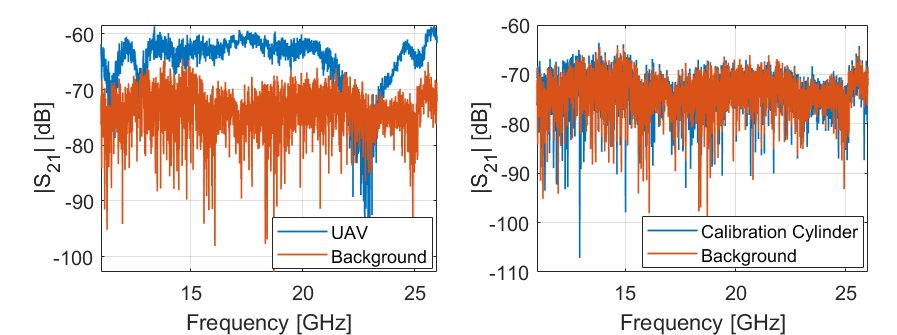}
\caption{Representative frequency-domain responses of the DJI Phantom~2,
calibration cylinder, and background. The background can mask the target echo,
which motivates coherent subtraction.}
\label{fig:frequency_domain_response}
\end{figure}

\begin{figure}[!t]
\centering
\includegraphics[width=\columnwidth]{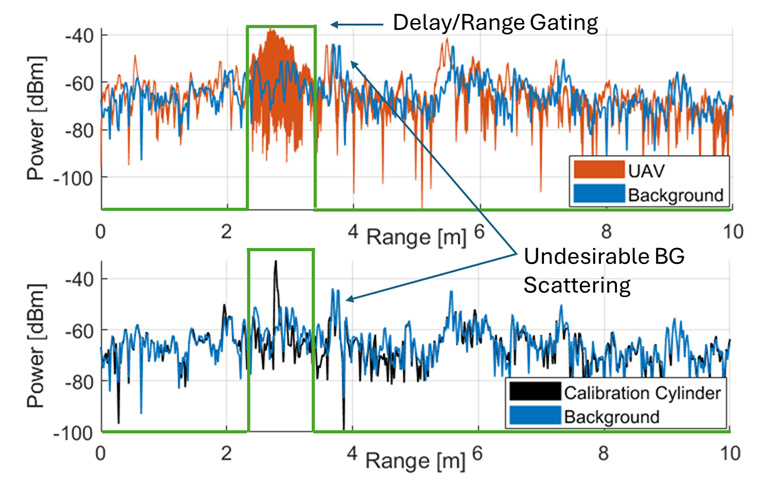}
\caption{Representative range-domain responses after coherent background
subtraction. Range gating isolates the target-dominated response and suppresses
residual environmental multipath.}
\label{fig:comst_range_gating}
\end{figure}

Fig.~\ref{fig:frequency_domain_response} illustrates why coherent background
subtraction is required: the environmental response can partially mask the UAV
return. Fig.~\ref{fig:comst_range_gating} then shows how the delay-domain gate
isolates the target-dominated response. Because the target and reference are
measured at the same location with the same instruments and antenna geometry,
their ratio removes common gain and propagation terms to first order. It does
not remove every uncertainty; alignment, gate selection, reference-target
accuracy, near-field behavior, residual background variation, and polarization
remain relevant. The final data are indexed by frequency, azimuth, UAV type,
and, for the Phantom~2, bistatic angle.

\subsubsection{Measurement Results}

The measurements show that UAV RCS depends jointly on frequency, aspect,
platform, and observation geometry. For the measured UAVs and geometries,
monostatic RCS is generally higher than the corresponding bistatic RCS.
This observation reflects the sampled scattering directions rather than a
universal ordering between monostatic and bistatic responses. The mean RCS also tends to increase as the
target becomes electrically larger at higher frequencies. Nevertheless, the
frequency response is not smooth: local peaks and notches occur because
different structures and components of the sensing target dominate at
different wavelengths and viewing angles.

\begin{figure}[!t]
\centering
\includegraphics[width=\columnwidth]{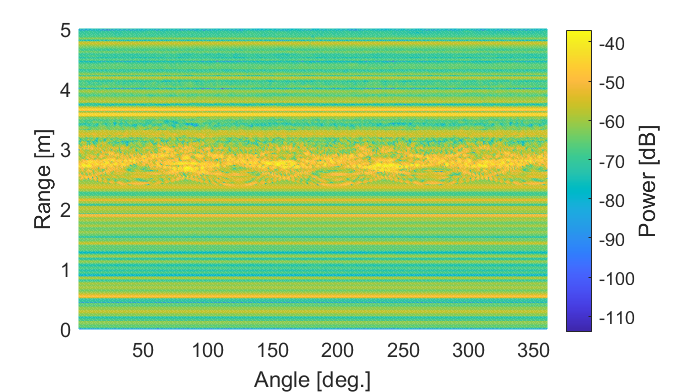}
\caption{Measured range--azimuth profile of the DJI Phantom~2. The signature
appears near the expected range and varies rapidly with azimuth because of
aspect-dependent scattering.}
\label{fig:range_azimuth_profile}
\end{figure}

\begin{figure}[!t]
\captionsetup[subfloat]{font=footnotesize,labelfont=footnotesize}
\centering
\subfloat[]{\includegraphics[width=0.49\columnwidth]{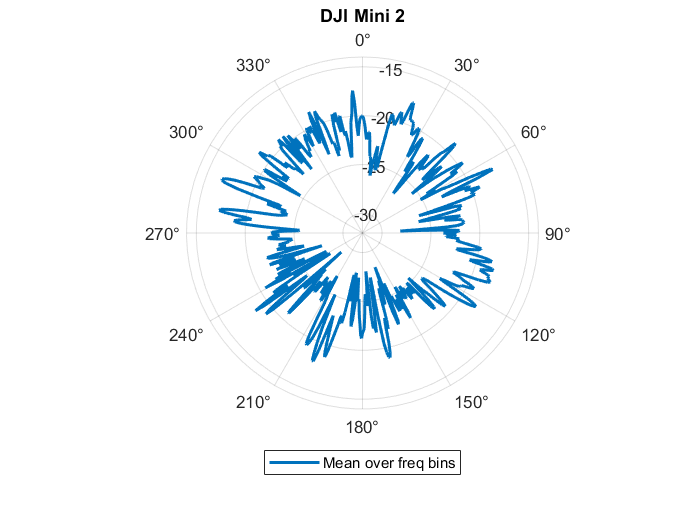}}
\hfill
\subfloat[]{\includegraphics[width=0.49\columnwidth]{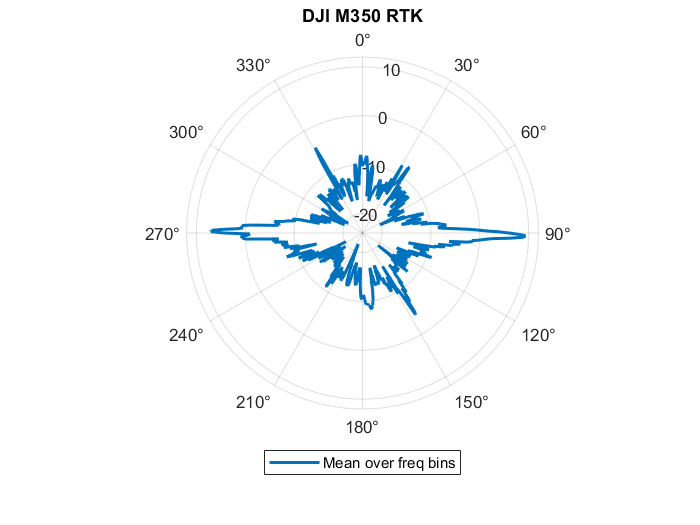}}\\
\subfloat[]{\includegraphics[width=0.49\columnwidth]{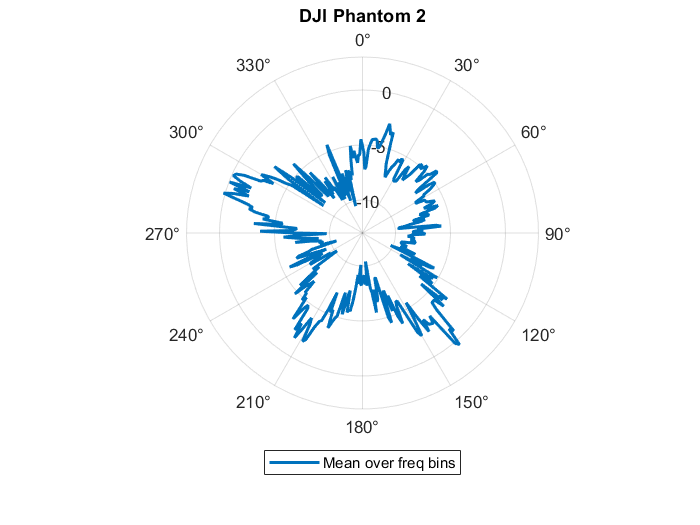}}
\hfill
\subfloat[]{\includegraphics[width=0.49\columnwidth]{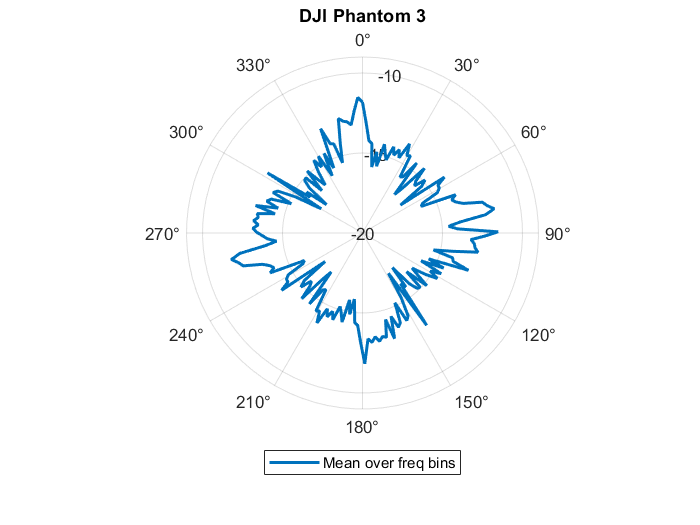}}
\caption{Representative measured wideband RCS patterns of four UAV data
sets. The angular fluctuations motivate an aspect-dependent target-visibility
model rather than a fixed RCS value.}
\label{fig:comst_rcs_patterns}
\end{figure}

A second observation is the pronounced angular variability visible in
Figs.~\ref{fig:range_azimuth_profile} and~\ref{fig:comst_rcs_patterns}.
Contributions from the fuselage, arms, landing structures, and rotor assemblies
combine constructively at some aspects and destructively at others, producing
rapid RCS fluctuations with observation angle. Consequently, a UAV that is
strongly visible from one geometry may become substantially weaker after only
a small change in aspect.

Across the evaluated data sets, the frequency-averaged azimuthal RCS can be
represented compactly by a log-normal distribution. Its fitted parameters,
however, remain platform- and band-dependent. A single omnidirectional RCS
constant therefore cannot reproduce the behavior illustrated in
Fig.~\ref{fig:comst_rcs_patterns}.

\begin{figure}[!t]
\centering
\includegraphics[width=0.84\columnwidth]{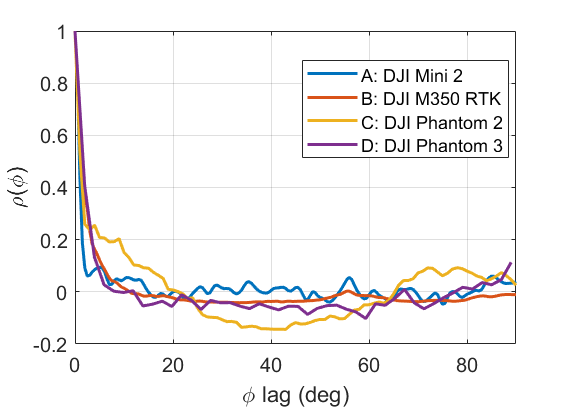}
\caption{Measured angular autocorrelation of UAV RCS. The rapid decay over a
few degrees indicates narrow angular coherence and strong aspect dependence.}
\label{fig:comst_rcs_acf}
\end{figure}

The fitted distribution describes the marginal RCS statistics but does not
specify how rapidly the target response changes with angle.
Fig.~\ref{fig:comst_rcs_acf} therefore provides a complementary second-order
description through the normalized angular autocorrelation function. The
angular autocorrelation falls rapidly over only a few degrees for all four UAV
data sets. In a moving scenario, this spatial characteristic becomes temporal
fluctuation through target attitude and transmitter--target--receiver geometry.
Independent RCS draws at every sample would discard the measured coherence,
whereas one static draw would overstate its persistence.

\begin{figure}[!t]
\centering
\includegraphics[width=0.82\columnwidth]{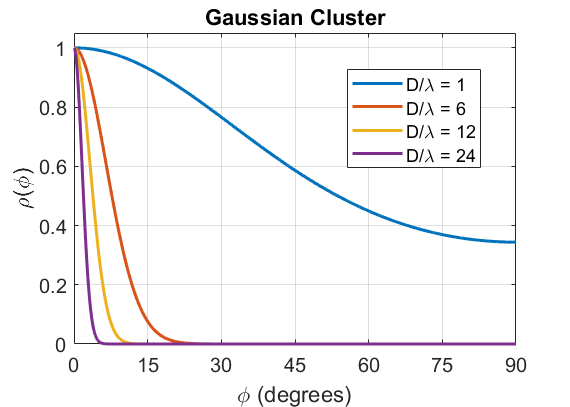}
\caption{Gaussian-cluster interpretation of angular RCS decorrelation.
Electrically larger targets decorrelate more rapidly with angle, consistent
with the measured UAV behavior.}
\label{fig:gaussian_cluster_interpretation}
\end{figure}

The Gaussian-cluster interpretation of a sensing target in
Fig.~\ref{fig:gaussian_cluster_interpretation} provides a compact physical
explanation for this behavior. Contributions from an electrically large set of
scattering components combine into angular lobes and notches as their relative
phases change. The resulting rapid decorrelation is therefore consistent with
the electrically large character of the UAVs over the considered frequency
bands.

\begin{figure*}[!t]
\captionsetup[subfloat]{font=footnotesize,labelfont=footnotesize}
\centering
\subfloat[]{\includegraphics[width=0.49\textwidth]
{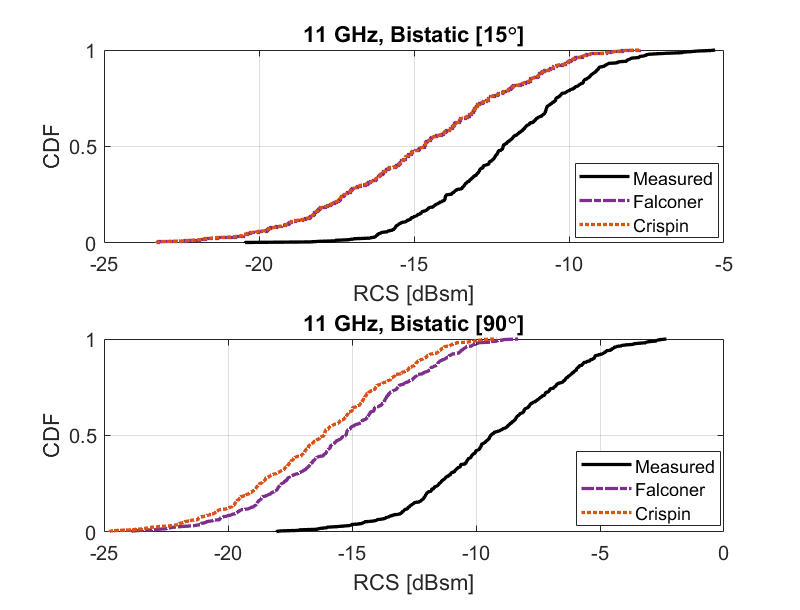}}
\hfill
\subfloat[]{\includegraphics[width=0.49\textwidth]
{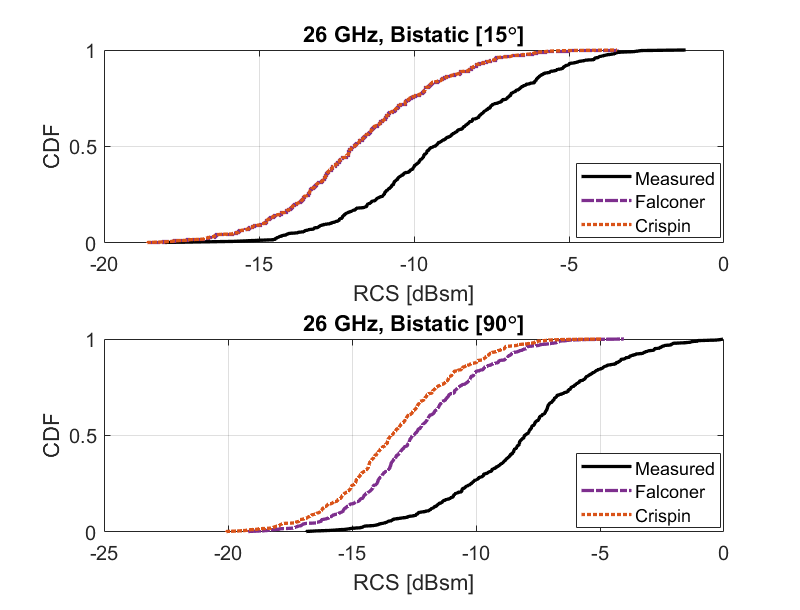}}
\caption{Measured bistatic UAV RCS distributions and predictions from the
Crispin and Falconer monostatic--bistatic equivalence models at (a) 11~GHz and
(b) 26~GHz. Each panel compares bistatic angles of $15^\circ$ and $90^\circ$.
The mismatch increases for some wideband, large-angle conditions.}
\label{fig:mbet_comparison}
\end{figure*}

The availability of bistatic measurements also enables a separate question to
be examined: whether the bistatic target response can be inferred reliably
from monostatic data. Classical monostatic-to-bistatic equivalence
approximations, including the Falconer and Crispin formulations, can reproduce
broad RCS trends at small bistatic angles~\cite{falconer1988near,4415533}. The
Phantom~2 data in Fig.~\ref{fig:mbet_comparison} show that agreement between classical monostatic-to-bistatic equivalence approximations can degrade at larger angles and for some frequency conditions. Shadowing among target parts, multiple scattering, structural periodicity, and wideband response can violate the smooth-target assumptions behind most equivalence theorems. Direct bistatic measurements, or a stochastic correction calibrated against such measurements, are therefore preferable for distributed UAV sensing when these assumptions are not satisfied.

\subsubsection{Implications for 3GPP-Oriented ISAC Evaluation}

The measurements lead to several implications for 3GPP-oriented ISAC channel
modeling and evaluation. First, UAV RCS should depend on target class and
orientation relative to the sensing nodes. A single fixed value cannot
adequately represent sensing reliability, tracking stability, or distributed
sensing performance. Moreover, angular visibility should be modeled as a
correlated stochastic process: a geometrically valid target path can still
become weak when the transmitter--target--receiver geometry coincides with an
RCS notch. Second-order descriptors such as angular level-crossing rate and
notch width can compactly characterize the frequency and angular extent of such
low-visibility regions.

\begin{figure}[!t]
\centering
\includegraphics[width=0.80\columnwidth]{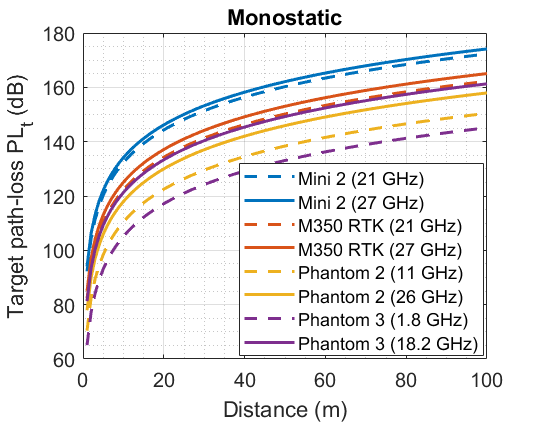}
\caption{RCS-based target-induced path-loss modeling for monostatic sensing,
connecting measured UAV scattering with ISAC performance evaluation.}
\label{fig:target_pathloss_mono}
\end{figure}

\begin{figure}[!t]
\centering
\includegraphics[width=0.80\columnwidth]{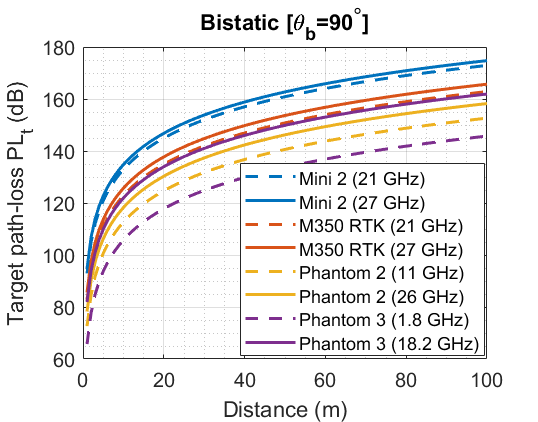}
\caption{RCS-based target-induced path-loss modeling for bistatic sensing.
Mean loss depends strongly on frequency and platform, while angular variation
remains important for sensing reliability.}
\label{fig:target_pathloss_bistatic}
\end{figure}

The fitted RCS statistics can also be incorporated into target-induced
path-loss models, as illustrated in Figs.~\ref{fig:target_pathloss_mono}
and~\ref{fig:target_pathloss_bistatic}. In the measured data, UAV class and
frequency strongly influence the average target-induced loss, whereas
instantaneous bistatic visibility retains a pronounced angular dependence.
Consequently, models intended for coverage-level evaluation may rely on
averaged statistics, while detection, tracking, and multi-node sensing
evaluations should retain the angular fluctuations.

Finally, the bistatic measurements indicate that monostatic-to-bistatic
equivalence should be used cautiously. Approximate transformations may remain
useful in restricted geometries, but direct bistatic measurements or
measurement-calibrated stochastic models are preferable when target complexity
violates their underlying assumptions.

These observations can be translated into the TR~38.901 target-channel
abstraction by assigning each target class a frequency-dependent RCS
distribution together with an angular correlation model. When target
orientation is available, a directional response can represent large-scale
visibility. Otherwise, a correlated stochastic process can reproduce
orientation-dependent fluctuations without treating successive observations
as independent. The bistatic correction should likewise depend on the
transmit and receive directions~\cite{ref11,unified_rcs_3gpp_isac}.

The present campaign does not characterize every parameter required by a
complete target-channel model. The fixed antenna polarization prevents
estimation of the full polarimetric target-transfer matrix and cross-polarized
response. Because the UAVs remain stationary on a turntable, the measurements
do not capture rotor micro-Doppler, translational motion, or attitude dynamics.
Elevation dependence is also not systematically evaluated, and the Phantom~2
measurements rely on a controlled, non-anechoic near-field setup. These
limitations define the scope within which the extracted RCS statistics should
be interpreted and motivate the broader measurement challenges discussed in
the subsequent section.

These measurements provide target-response inputs for the broader validation
roadmap developed in Section~\ref{sec:challenges}-F.


%% file: sections/07_challenges.tex
\section{Open Challenges and Path Toward Deployment}
\label{sec:challenges}

This section identifies five coupled technical challenges for cellular ISAC
and presents a validation roadmap toward network deployment.

\subsection{Channel and Target-Response Generalization}

The remaining challenge is transferring target-response models beyond the
platforms and operating conditions from which they were calibrated.
Section~VI-C showed how measured UAV responses can parameterize a target
channel within a limited range of frequencies, polarizations, and observation
geometries. Broader deployment requires models that remain valid across target
classes, structural variants, motion states, sensing modes, and environmental
conditions. The TR~38.901 framework provides the target-channel structure, but
it does not provide universal parameters for these combinations
~\cite{ref11,tr38765}.

Future measurement programs should cover frequency, azimuth, elevation,
bistatic angle, co- and cross-polarized response, body and rotor motion,
blockage, and environmental clutter. They should retain calibration uncertainty and second-order properties such as angular and temporal correlation, visibility duration, and cross-link correlation. Repeated trajectories are also needed to distinguish persistent target characteristics from measurement noise and environmental variation.

A promising research direction is a hierarchical target-response model. A
class-level distribution can represent variation across a target population,
while a platform-specific state captures persistent differences in size,
shape, material composition, and structural configuration. Correlated
orientation- and motion-dependent processes can then represent local
visibility and micro-Doppler. Deterministic electromagnetic or ray-based
responses can characterize specific objects, while stochastic residual models
enable system-level scaling. The model should also quantify uncertainty so
that detectors and trackers do not treat a nominal target coefficient as an
exact physical value
~\cite{wei2025isacchannel,unified_rcs_3gpp_isac,isac_channel_model_3gpp}.

\begin{figure}[!t]
\centering
\includegraphics[width=0.9\columnwidth]
{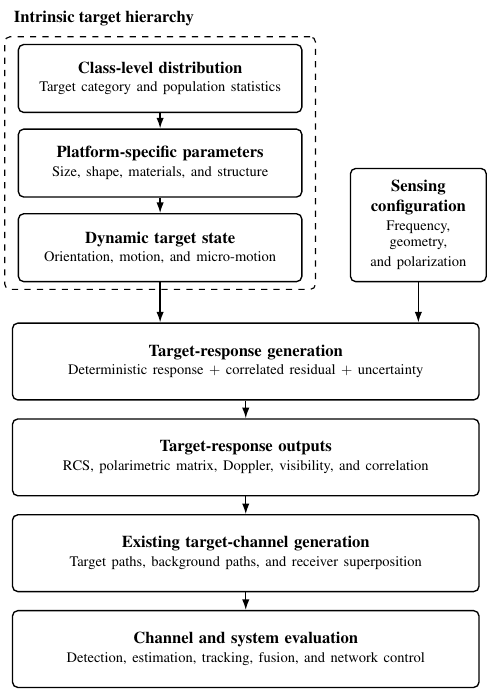}
\caption{Conceptual hierarchical target-response model. Class- and
platform-level target properties, dynamic target state, and sensing
configuration determine the target-response parameters supplied to the
existing target-channel generation framework.}
\label{fig:hierarchical_target_response_model}
\end{figure}

\subsection{Multi-Node Coordination, Fusion, and Tracking}

Multi-TRP sensing promises broader spatial coverage, complementary observation
geometries, improved target visibility, and greater resilience to blockage.
Realizing these benefits, however, requires observations from different TRPs
to maintain a consistent physical interpretation. Nodes may differ in target
visibility, clutter conditions, oscillator error, clock bias, coordinate
accuracy, update rate, detector threshold, and transport delay. Fusion can
perform worse than a single reliable link when measurements are incorrectly
associated or correlated errors are treated as independent
~\cite{li_2024_twc,meng_2025_wc,meng_2025_twc}.

The remaining standardization gap concerns how sensing reports from multiple
TRPs are interpreted, associated, and fused after they have been transported
to a common processing function. The emerging Ns/SEN2 framework can support
sensing control and the transfer of Level~C or Level~D information, but
interoperable multi-node operation also requires common timing and coordinate
references, measurement identifiers, association and object-lifecycle rules,
confidence definitions, result ownership, missing-report handling, and failure
recovery. Tracking further couples radio scheduling to state uncertainty
because a link that provides little information at the current target position
may become geometrically valuable as the target moves.

Evaluation of multi-node fusion should begin with Monte Carlo simulation that
represents correlated target visibility, clutter, report loss, and transport
delay. The simulation should retain the target-present and target-absent
denominators required to estimate low FAPs. Experimental validation can then
progress from cabled or common-clock multi-node setups to over-the-air
synchronization and trials with independently operated sites. Each stage
should record timing residuals, association errors, report latency, dropped
measurements, and false alarms. A promising research direction is
uncertainty-aware asynchronous fusion that selects the report level and
participating links according to the observation geometry, transport capacity,
and track state, rather than assuming that every observation is available and
equally reliable.

\subsection{Resource Allocation, Waveforms, and Interference Management}

Sensing and communication must share finite time, frequency, spatial, power,
and antenna resources while meeting the requirements of both functions.
Sensing generally benefits from wider bandwidth, longer coherent observation,
larger array aperture, greater illumination power, and repeated observations.
Communication places greater emphasis on low latency, reliable control,
scheduling flexibility, and energy efficiency. Resource allocation must
therefore balance sensing resolution and reliability against communication
quality of service and implementation cost. A promising direction is
sensing-lifecycle-aware allocation. Initial detection can use broad spatial
coverage and sufficient coherent resources, whereas tracking can use
prediction-guided beams and less frequent or narrower-band updates
~\cite{zong_2025_tvt,yang_2024_twc}.

Waveforms and reference signals should be designed jointly with their resource
mapping and intended sensing tasks. A sensing-optimized signal may provide
favorable range--velocity ambiguity characteristics but have a high
PAPR, limited communication reuse, or excessive
resource overhead. Conversely, an existing communication reference signal may
not provide sufficient bandwidth, coherent duration, or sampling density for
the required sensing performance. Standardization must therefore consider
reference-signal patterns, multiplexing methods, and resource configurations
as parts of a coupled radio design
~\cite{wei_2023_iotj,niu_2025_iotj}.

Interference and beam management introduce additional requirements.
Monostatic self-interference, mutual sensing interference, and
sensing--communication cross-interference have different characteristics and
require corresponding measurement and mitigation methods. Target-aware beam
selection also differs from communication beam selection because the strongest
data beam may not provide the most effective target illumination or bistatic
observation geometry.

Alternative designs should be compared using the same total time, bandwidth,
power, and antenna resources. Link-level evaluation should include phase
noise, nonlinearities, receiver dynamic range, synchronization error, and
structured interference. System-level evaluation should then propagate the
calibrated detection and decoding behavior into scheduling, coverage, energy
consumption, and mobility key performance indicators. Communication
reliability constraints should be preserved so that an apparent sensing gain
does not result from an increased resource budget.

\subsection{RAN and Core-Network Evolution}

The Release~20 RAN and service architectures reviewed in
Sections~\ref{subsubsec:ran_architecture}
and~\ref{subsubsec:service_architecture} establish an initial control and
reporting path for gNB-based sensing. The remaining challenge is to extend this constrained baseline into a scalable sensing service that maintains consistent state as sensing entities, target areas, processing locations, and serving network functions change~\cite{tr2370014,ts23137,tr2380101}.

At the RAN level, the gNB-DU, gNB-CU, and SenF require unambiguous ownership of measurement configuration, processing state, report identity, and failure
recovery. Time-critical radio processing should remain close to the DU, while
the CU and SenF coordinate longer-term association, fusion, and service
control. At the service level, broader operation requires sensing-entity
discovery and reselection, context transfer, result ownership, storage and
reuse policies, and continuity across multiple access nodes or SenF instances.
These functions must prevent duplicated tasks, inconsistent tracks, and loss
of sensing context during mobility, failure, or load redistribution.

Validation requires an interface-in-the-loop implementation rather than
message-format inspection alone. Tests should vary report size, update rate,
packet loss, reordering, processing load, transport latency, node restart, and
partial sensing-entity failure. End-to-end trials should trace a request from
application authorization through sensing-entity selection, RAN configuration,
report generation, fusion, and notification. A promising direction is a
hierarchical SenF realization in which radio-proximate functions perform
time-critical preprocessing, regional functions coordinate association and
fusion, and service-level functions manage authorization, policy, and result
delivery. The functional split should be determined by latency, transport,
privacy, scalability, and failure-containment requirements.

\begin{table*}[!t]
\caption{Standards-to-deployment roadmap for cellular ISAC evaluation}
\label{tab:validation_roadmap}
\centering
\scriptsize
\setlength{\tabcolsep}{3.5pt}
\begin{tabular}{P{0.08\textwidth} P{0.20\textwidth} P{0.40\textwidth} P{0.24\textwidth}}
\toprule
Stage & Principal question & Reproducibility requirements & Evaluation scope and uncertainty \\
\midrule
Standardized scenario & Is the use case and KPI defined consistently? & Document version, clause/table mapping, geometry, target class, sensing mode, KPI definition, and separately listed study-specific choices & Supports assumption alignment and scenario comparison \\
Calibrated channel/link model & Does the implementation reproduce the intended physical and link response? & Source/configuration version, fixed seeds, array/waveform details, calibration cases, reference curves or measurements, thresholds, and acceptance gates & State calibration range, residual model error, and parameters not covered by the reference \\
System-level simulation & Does the mechanism improve network KPIs at scale? & Paired drops, spatial consistency, interference and mobility models, link-to-system mapping, procedure state machine, raw events, denominators, and inference code & Report conditioning, rare-event resolution, multiplicity, sensitivity cases, and abstraction limits \\
Real-time prototype & Can the waveform, processing, and control loop operate within timing and hardware limits? & Hardware/software versions, clocking, RF settings, resource allocation, processing split, latency traces, intermediate outputs, and failure logs & State tested geometry, target, load, dynamic range, and repeatability; feasibility is not network-scale reliability \\
Controlled measurement & Are channel, clutter, and target assumptions physically supported? & Calibrated geometry, reference object or transfer standard, raw and processed responses, background data, uncertainty budget, polarization, target state, and ground truth & State near-/far-field condition, environment, sampled dimensions, and limits of transfer across bands and targets \\
Field trial & Does the integrated service remain useful in its deployment environment? & Deployment map, environmental context, traffic/load, ground truth, end-to-end logs, failures, data-governance protocol, and repeat campaigns & Report coverage of operating conditions, missing data, safety/privacy constraints, and external validity \\
\bottomrule
\end{tabular}
\end{table*}

\subsection{Security, Privacy, and Trustworthy Sensing}

Cellular sensing can observe people, vehicles, property, and activity even when the target is not a network subscriber
~\cite{tr_22_837,etsi_gr_isc_004}. The resulting security and privacy
challenge therefore extends beyond protecting communication payloads. A
legitimate sensing service can reveal sensitive location or behavioral
information, while a malicious actor can spoof a target, replay measurements,
manipulate time or position references, corrupt a track, or infer information
from delivered sensing results
~\cite{etsi_gr_isc_004,martins_2025_ojcoms}. Within 3GPP, the Release~20 SA3 study in draft TR~33.777 examines the security and privacy aspects of ISAC~\cite{tr33777}.

The service requirements and SA2 architecture introduce authorization and
controlled delivery to trusted and untrusted application functions. Trustworthy sensing also requires purpose limitation, geographic and temporal restrictions, retention policies, consent where applicable, data provenance, auditing, and revocation~\cite{tr_22_837,tr2370014,ts23137}. A confidence value is not a security guarantee because a precisely formatted, high-confidence report may still originate from a compromised sensing entity or a biased fusion process. Similarly, transferring only object-level reports does not ensure privacy. An object trajectory may reveal identity, behavioral patterns, frequently visited locations, or other sensitive information. Protection must therefore cover raw measurements, intermediate processing outputs, and derived sensing results throughout their collection, processing, storage, and delivery.

Security validation should include explicit threat models and adversarial
testing at every sensing-report level
~\cite{etsi_gr_isc_004,martins_2025_ojcoms}. Experiments should quantify both
detection performance and network-control consequences under spoofing,
jamming, replay, timestamp manipulation, compromised sensing entities, and
malformed or selectively missing reports
~\cite{yildirim_2025_jiot,bartoletti_2026_jsac}. Promising research directions
include authenticated measurements, secure time and location references,
traceable fusion, cross-link plausibility checks, privacy-aware aggregation,
and policy enforcement at the SenF and service-delivery interface
~\cite{etsi_gr_isc_004,martins_2025_ojcoms}. Security performance should be
reported together with sensing utility and latency because stronger protection
may affect the timeliness of network decisions.

\subsection{Reproducible Evaluation Roadmap}

Table~\ref{tab:validation_roadmap} organizes cellular ISAC validation into six
stages that progressively increase model scope, implementation detail, and
operational realism. The stages do not prescribe a fixed sequence.
Measurements can inform channel and link calibration before large-scale
simulation, while prototype and field results can expose assumptions that
require earlier models to be revised. The table instead identifies the
validation objective, reproducibility materials, and interpretation limits
associated with each stage~\cite{sargent_2013_validation}.

Traceability should be maintained throughout this process. Standards,
software, configurations, data, and figures should use stable identifiers and
versions. Raw events and their denominators should be retained, including
failures and missing observations. Calibration, tuning, and final evaluation
data should be separated. Uncertainty and latency should also be propagated
across interfaces rather than replaced by perfect intermediate state
information~\cite{peng_2011_reproducible,wilkinson_2016_fair}.

The roadmap distinguishes research validation from standards conformance. A
standards-grounded simulation evaluates a research hypothesis under traceable
assumptions. Calibration determines whether a model reproduces a defined
reference, while prototype and field evaluations examine implementation
behavior under stated conditions. Conformance can be established only through
a normative test procedure with specified configurations, measurements, and
acceptance criteria~\cite{3gpp_ts_38141_1,3gpp_ts_38521_1}.


%% file: sections/08_conclusion.tex
\section{Conclusion}
\label{sec:conclusion}

In this article, we connected three views of cellular ISAC that are often
evaluated separately: 3GPP standardization, reproducible performance
evaluation, and measurement-based target characterization. We first traced the progression from Release-19 service requirements and target-aware channel modeling to the Release-20 NR sensing evaluation framework, RAN-side SenF--gNB procedures, and SA2 service architecture. The current baseline remains focused on gNB-based UAV sensing, while broader target classes, UE participation, multistatic cooperation, and network-wide sensing control remain study and evolution directions toward 6G.

We then used two simulation case studies to show why the evaluation level and operating point matter. In the multi-TRP UAV case study, policies frozen at a common calibration false-alarm budget showed that hybrid fusion improves detection while retaining conditional estimation accuracy close to hard-decision fusion. The independently generated extension supported this detection improvement, while sparse false-alarm events limited conclusions about false-alarm superiority. In the sensing-assisted predictive-handover case study, practical sensing assistance reduced mean handover-attributable interruption from 95.3 to 58.0~ms and increased mean user throughput from 14.50 to 15.79~Mbit/s relative to conventional handover. However, the sparse primary handover failures do not support a general reliability-superiority claim. Instead, the results show the importance of sensing age and uncertainty, common execution criteria, and separate control- and data-plane link adaptation.

Finally, the experimental case study provided measurement-based characterization supporting target modeling. Monostatic and bistatic UAV measurements showed that RCS varies with frequency, platform type, and observation aspect, with correlation over only a limited angular interval. These results support modeling target visibility as a correlated process rather than as a fixed scattering coefficient. The campaign establishes a calibrated baseline under fixed polarization and controlled turntable rotation, while indicating how measurements across polarization, elevation, bistatic angle, and target motion can extend the model toward a dynamic polarimetric target response.

Together, these results provide a standards-to-deployment perspective for
interpreting cellular ISAC performance according to the adopted model
resolution, operating conditions, and validation level. This perspective
supports the progression from isolated sensing demonstrations toward reliable cellular ISAC operation.